\documentclass[aps,prb,a4paper,twocolumn,showkeys,preprintnumbers,longbibliography]{revtex4-1}

\usepackage[pdftex]{graphicx}
\DeclareGraphicsExtensions{.png, .jpg, .pdf}

\graphicspath{{Figures/PNG/}{Figures/PDF/}{Figures/}}

\usepackage{longtable}
\usepackage{dcolumn} % Align table columns on decimal point
\usepackage{bm} % bold math
\usepackage{amsmath} %better math support (more symbols, and other)
\usepackage{amssymb} % Adds additional AMS symbols.
\usepackage{subfigure}

    \usepackage[ pdftex, plainpages = false, pdfpagelabels,
                 pdfpagelayout = OneColumn, % display single page, advancing flips the page - Sasa Tomic
                 bookmarks,
                 bookmarksopen = true,
                 bookmarksnumbered = true,
                 breaklinks = true,
                 linktocpage,
                 pagebackref,
                 colorlinks = true,
                 linkcolor = blue,
                 urlcolor  = blue,
                 citecolor = red,
                 anchorcolor = green,
                 hyperindex = true,
                 hyperfigures
                 ]{hyperref}
\begin{document}

%\preprint{APS/123-QED}

% The following information is for internal review, please remove them for submission
%\widetext
%\leftline{ \begin{scriptsize} Version as of \today \end{scriptsize} }
%\leftline{Primary authors: Mohammed Moaied}
%\leftline{To be submitted to (PRL, PRD-RC, PRD, PLB; choose one.)}
%\leftline{Comment to {\tt moaied5@yahoo.com} by Mohammed Moaied, and Juan-Jos\'{e} Palacios}
%\centerline{\em D\O\ INTERNAL DOCUMENT -- NOT FOR PUBLIC DISTRIBUTION}

% the following line is for submission, including submission to the arXiv!!
%\hspace{5.2in} \mbox{Fermilab-Pub-04/xxx-E}

\title{\textbf{Hydrogenation-induced ferromagnetism on graphite surfaces}}

\author{
\firstname{Mohammed}
\surname{Moaied},
}
\email{moaied5@yahoo.com}

\author{
\firstname{J. V.}
\surname{\'Alvarez},
}
\email{jv.alvarez@uam.es}

\affiliation{Departamento de F\'{i}sica de la Materia Condensada, Universidad Aut\'{o}noma de Madrid, Cantoblanco, 28049 Madrid, Spain.  }

\author{
\firstname{J. J.}
\surname{Palacios}
}
\email{juanjose.palacios@uam.es}
%\homepage{https://sites.google.com/site/palaciosjuanjose/home}

\affiliation{Departamento de F\'{i}sica de la Materia Condensada, Instituto Nicol\'as Cabrera (INC), and Condensed Matter 
Physics Center (IFIMAC), Universidad Aut\'{o}noma de Madrid, Cantoblanco, 28049 Madrid, Spain.
}

%\thanks{Work supported by Spanish Ministry of Education}

\date{\today}

% ------------------------------------------------------------------------------
% -------------------------- abstract ------------------------------------------
% ------------------------------------------------------------------------------

\begin{abstract}
We calculate the electronic structure and magnetic properties of hydrogenated graphite surfaces using van der Waals density functional theory (DFT) and
model Hamiltonians.
We find, as previously reported, that the interaction between hydrogen atoms on graphene favors adsorption on different
sublattices along with an antiferromagnetic coupling of the induced magnetic moments. On the contrary, when hydrogenation takes place
on the surface of graphene multilayers or graphite (Bernal stacking), the interaction between hydrogen atoms competes with the different
adsorption energies of the two sublattices. This competition may result in all hydrogen atoms adsorbed on the same sublattice and, thereby,  in a ferromagnetic state for low concentrations. Based on the exchange couplings
obtained from the DFT calculations, we have also evaluated the Curie temperature by mapping this system onto an Ising-like model with randomly located spins.
Remarkably, the long-range nature of the magnetic coupling in these systems makes the Curie temperature size dependent
and larger than room temperature for typical concentrations and sizes.
\end{abstract}


%\pacs{Valid PACS appear here}% PACS, the Physics and Astronomy Classification Scheme.

\maketitle

% --------------------------------------------------------------------------------
% -------------------------- Introduction ----------------------------------------
% --------------------------------------------------------------------------------

\section{Introduction}

Hydrogenation of carbon nanostructures is recently attracting a lot of interest as a methodology that allows for the tuning of their mechanical, electronic, and magnetic properties.
In contrast to direct manipulation of the carbon atoms, e.g., creating vacancies \cite{Gomez-Navarro2005,PhysRevLett.104.096804} or reshaping edges \cite{Jia27032009}, hydrogenation can effectively affect
the electronic properties in a similar manner with the advantage that is a reversible process. For instance,
hydrogenation of graphene was found, both theoretically and experimentally, to be a way to turn graphene from a gapless semiconductor into a gapful one with a tunable band gap \citep{Elias30012009, Sessi2009, Haberer2010, Yang2010, Balog2010}. It has also been predicted that partial hydrogenation may induce interesting magnetic properties in graphene with potential applications in spintronics \citep{PhysRevB.81.165409}. For instance, H-induced ferromagnetism is expected under some very particular conditions \cite{Zhou2009}. Recent experiments on hydrogenated or fluorinated graphene, however, do not show evidence of ferromagnetism, but rather of paramagnetic behaviour \cite{PhysRevLett.105.207205,Nair12,McCreary12}.

Many calculations related to the adsorption of H atoms on graphene have been reported in the literature, 
mostly being based on first-principles or density functional theory (DFT). All the reports coincide in that 
adsorptive carbon atoms are puckered and, most importantly, that the covalent bond between carbon and H leads 
to magnetic moments on neighboring carbon atoms totalling 1.0 
$ \mu_{{\rm B}}$\citep{PhysRevLett.91.017202,PhysRevB.77.035427, Casolo2009,  PhysRevLett.107.016602}. 
Such spin polarization is mainly localized around the adsorptive carbon atom.
The magnetic coupling between H atoms adsorbed on graphene has also been studied and 
basically follows the rules expected from Lieb's theorem \cite{PhysRevLett.62.1201}.
Graphene is a single layer of carbon atoms bonded together in a bipartite honeycomb structure. It is thus formed by two interpenetrating triangular sublattices, A and B,
such that the nearest neighbors of an atom A belong to the sublattice B and vice versa \citep{saito1998physical}.
Three different magnetic states can be triggered by a H pair, namely, non-magnetic, ferromagnetic, and antiferromagnetic.
The most energetically stable configuration corresponds to
having both H atoms adsorbed on two nearest-neighbor carbon atoms, leading to a non-magnetic 
ground state \citep{PhysRevB.77.035427, Casolo2009, PhysRevB.81.165409, PhysRevLett.107.016602}.
When both H atoms are on the same sublattice they are coupled ferromagnetically with total spin $S=1$. 
When the pair of H atoms is adsorbed on
different sublattices, but sufficiently far away from each other, they induce magnetic moments that 
couple antiferromagnetically ($S=0$) \cite{PhysRevB.81.165409}. As we show here, for similar distances between the 
H atoms the ferromagnetic coupling is always favored over
the antiferromagnetic one. Previous calculations for vacancy-induced magnetism in graphene
have shown similar results as long as the vacancies do not reconstruct \citep{PhysRevB.75.125408, PhysRevB.77.195428}.

Concerning graphite,  a few experimental studies, not free from controversy, have reported changes in 
the magnetic properties produced by irradiation of the graphite sample. 
The results show that graphite can become ferromagnetic at room temperature out of an originally non-magnetic sample. 
The ferromagnetic state appears at low concentration of the impurities induced by the 
irradiation and is independent of the irradiation ion type used \citep{PhysRevLett.91.227201, PhysRevB.81.214404}.
Unlike the case of graphene, not many theoretical studies have been reported in the literature
on the magnetic properties of irradiated or hydrogenated graphite.
Yazyev\citep{PhysRevLett.101.037203}, for instance, has studied the magnetic properties of 
hydrogenated graphite using a combination of mean-field Hubbard model and first-principles calculations.
He obtained, as expected, that the sublattices are inequivalent (approx. 0.16 eV) for hydrogenation in bulk.
Graphite is a semi-metal composed of stacked graphene layers. The typical Bernal stacking of these planes 
effectively breaks sublattice symmetry: A atoms (for instance) are located exactly above and below the atoms of
neighboring planes (\textbf{$ \alpha $} atoms from now on) while B atoms 
are located at the center of the hexagonal rings of the neighboring planes
(\textbf{$ \beta $} atoms)\citep{pauling1960nature}.

Here we are concerned with hydrogenation of the surface of graphite. First, through DFT calculations,
we revisit the energetics of a H pair
on graphene. We confirm previous results and, by considering very large supercells, 
we find the expected antiferromagnetic state when H atoms 
are adsorbed sufficiently far apart from each other on different sublattices. Next we present results for 
the adsorption energies on different sublattices for bilayer  and multilayer graphene.
Both sets of results are then combined to estimate the maximum average concentration 
for which all H atoms may occupy the same sublattice and, thereby, will be coupled ferromagnetically. 
We also compute the exchange coupling constants as a function of the relative distance between H atoms. 
Finally, we present a study of the Curie temperature in this system based on a 
Ising model constructed with the DFT coupling constants. Our results
support the possible existence of surface sublattice-polarized hydrogenation and concomitant ferromagnetism.

% --------------------------------------------------------------------------------
% -------------------------- Computational Method --------------------------------
% --------------------------------------------------------------------------------

% --------------------------------------------------------------------------------
% -------------------------- Results and Discussion ------------------------------
% --------------------------------------------------------------------------------

\section{Atomic, electronic, and magnetic structure of H atoms on graphene and graphene multilayers}

\subsection{Computational Details}

Our calculations are based on the DFT framework\citep{PhysRev.136.B864, PhysRev.140.A1133} 
as implemented in the SIESTA code
\citep{PhysRevB.53.R10441, 0953-8984-14-11-302}. We are mostly interested here in multilayer graphene and graphite where dispersion (van der Waals) forces
due to long-range electron correlation effects play a key role in the binding of the graphene layers. Therefore
we use the exchange and correlation nonlocal van der Waals density functional (vdW-DF)
of Dion et al. \citep{PhysRevLett.92.246401} as implemented by Rom\'{a}n-P\'{e}rez and Soler \citep{PhysRevLett.103.096102, PhysRevLett.103.096103}.
To describe the interaction between the valence and core electrons we used norm-conserved Troullier-Martins pseudopotentials \citep{PhysRevB.43.1993}.
To expand the wavefunctions
of the valence electrons a double-$ \zeta $ plus polarization (DZP) basis set was used \citep{PhysRevB.64.235111}. We experimented
with a variety of basis sets and
found that, for both graphene and graphite, the DZP produced high-quality results. The plane-wave cutoff energy for the wavefunctions was
set to 500 Ryd. For the Brillouin zone sampling we use $ 4\times4\times2 $ Monkhorst-Pack (MP) 
$k$-mesh for the largest $ 12\times12\times1 $ single-layer and for the bilayer graphene supercells.
We have also checked that the results are well converged with respect to the real space grid. 
Regarding the atomic structure,
the atoms are allowed to relax down to a force tolerance of 0.005 eV/\AA{}.
All supercells are large enough to ensure that the vacuum space is at least 25 \AA{}
so that the interaction between functionalized graphene layers
and their periodic images is safely avoided.
Spin polarization was included in the calculations because, as discussed in the introduction,
hydrogenation is known to induce magnetism in single-layer and, possibly, also in bilayer and multilayer graphene.

\subsection{Preliminary checks}

\begin{figure}[!htbp]
  \begin{center}
    \leavevmode
    \ifpdf
      \includegraphics[height=1.7in]{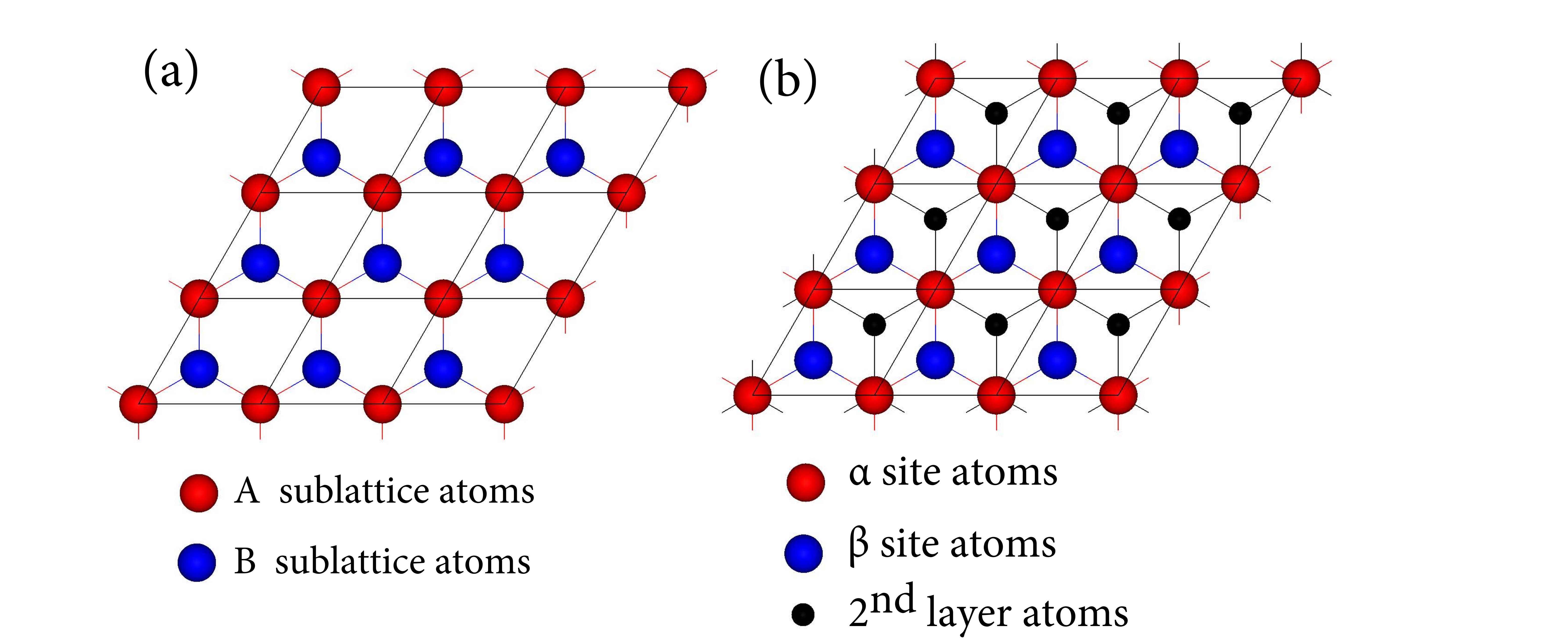}
    \else
      \includegraphics[bb = 92 86 545 742, height=3in]{structure-graphene-bilayer}
    \fi
    \caption{(Color online). Atomic structure of (a) single-layer and (b) bilayer graphene for a $ 3\times3\times1 $ supercell.}
    \label{structure-graphene-bilayer}
  \end{center}
\end{figure}

We begin our study by optimizing the geometric structures of the monolayer, bilayer graphene, and 
graphite in their natural nonmagnetic state.
The C-C bond lengths and cell parameters ($a$ and $c$) and the interlayer distances between the layers ($d$) 
 are listed in Table (\ref{table:structure}).
The accuracy of our procedure is very satisfactory when these magnitudes are contrasted against experimental values. For completeness, we present the atomic structures
of single-layer and bilayer graphene in Fig. \ref{structure-graphene-bilayer}. Different colors are used to stress different sublattices.

% -------------------------- Structure Table --------------------------------------
\begin{table}[ht]
\caption{Atomic structure parameters of monolayer, bilayer graphene, and graphite.}   % title of Table
\centering
\begin{center}

\begin{tabular}{  | p{2cm} | c | c | c | c | }                                 % centered columns (5 columns)
      \hline                                                                   % inserts single horizontal line
%heading
               & C-C bond (\AA{})     & $a$ (\AA{})    &$ c$ (\AA{})    &$ d$ (\AA{})    \\  % inserts table
    \hline%\hline                                                              % inserts double horizontal lines
    Graphene & 1.419 & 2.458 & 25       &     -      \\
    \hline
    Bilayer  & 1.420 & 2.459 & 25       & 3.353 \\
    \hline
    Graphite & 1.417 & 2.455 & 6.709 & 3.354 \\               % [1ex] adds vertical space
    Experimental &  & 2.456 \citep{sands1994introduction} & 6.696 \citep{sands1994introduction} &      \\
    \hline                                                                     %inserts single line
\end{tabular}

\end{center}
\label{table:structure}                                                        % is used to refer this table in the text
\end{table}
% ------------------------------------------------------------------------

% -------------------------- Structure ---------------------------------------

% ------------------------------------------------------------------------

\begin{figure}[!htbp]
  \begin{center}
    \leavevmode
    \ifpdf
      \includegraphics[height=3.4in]{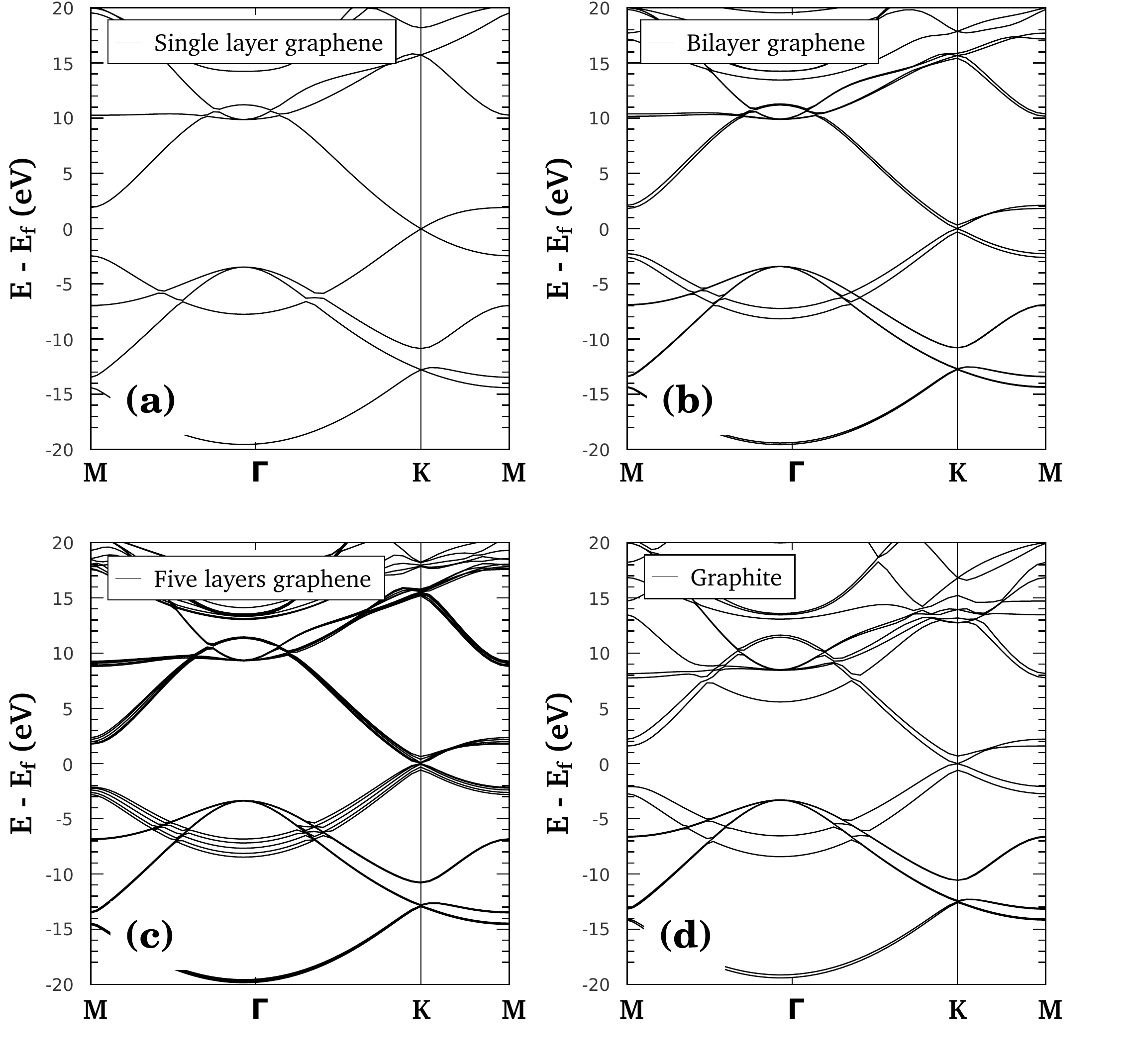}
    \else
      \includegraphics[bb = 92 86 545 742, height=3in]{Bands}
    \fi
    \caption{Electronic band structure of monolayer, bilayer, five-layer graphene, and graphite
calculated with a $ 100\times100\times2 $ MP-grid for a $ 1\times1\times1 $ supercell.}
    \label{Bands}
  \end{center}
\end{figure}

In Fig. \ref{Bands} 
we show the electronic band structure for monolayer, bilayer, five-layer graphene, and graphite
along the high-symmetry points M$\Gamma$KM. The well-known case of graphene is shown in Fig. \ref{Bands}(a), being the result similar to that found by many others (see, for instance
Ref. \citep{PhysRev.71.622, PhysRev.108.612, PhysRevB.66.035412}).
Since there are two basis atoms in graphene there is one pair of $\pi \pi^{\ast}$ bands of $p_{z}$ character,
which are degenerate at the K-point or Dirac Point, coinciding with the Fermi level.
 We have considered bilayer graphene in Bernal stacking, as for a typical
graphite arrangement. Since the basis consists now of four atoms, there are two pairs of $\pi \pi^{\ast}$ bands and there are four
sets of $p_{z}$-derived bands close to
the K-point as shown in Fig. \ref{Bands}(b). Due to the interaction between the graphene layers these bands split apart.
Consistent with previous theoretical works \citep{PhysRevLett.97.036803, PhysRevB.75.155115},
we find that, similar to  monolayer graphene, the bilayer graphene is also a zero-gap semiconductor with a pair of the $\pi \pi^{\ast}$
bands being degenerate at the K-point.  On the other hand, there is an energy gap of 0.8 eV between the other pair of
$\pi \pi^{\ast}$ bands. The band structure for five-layer graphene is shown in 
Fig. \ref{Bands}(c) which already anticipates the
characteristic band structure of graphite. For instance, at the $\Gamma$  point, five bands closely packed in energy 
manifest the emerging dispersion stemming from the perpendicular interlayer coupling.  
Finally, the bands of graphite are shown in Fig. \ref{Bands}(d). The results are also in agreement with 
previous works (see, e.g., Ref. \onlinecite{PhysRevB.25.4126}), exhibiting  a bandwith of 1.41 eV at the 
K-point\citep{PhysRevB.55.4999}.

% -------------------------------- Bands -------------------------------------

\subsection{Hydrogen atoms on monolayer graphene}

\subsubsection{One hydrogen atom}

\begin{figure}[!htbp]
  \begin{center}
    \leavevmode
    \ifpdf
      \includegraphics[height=1.4in]{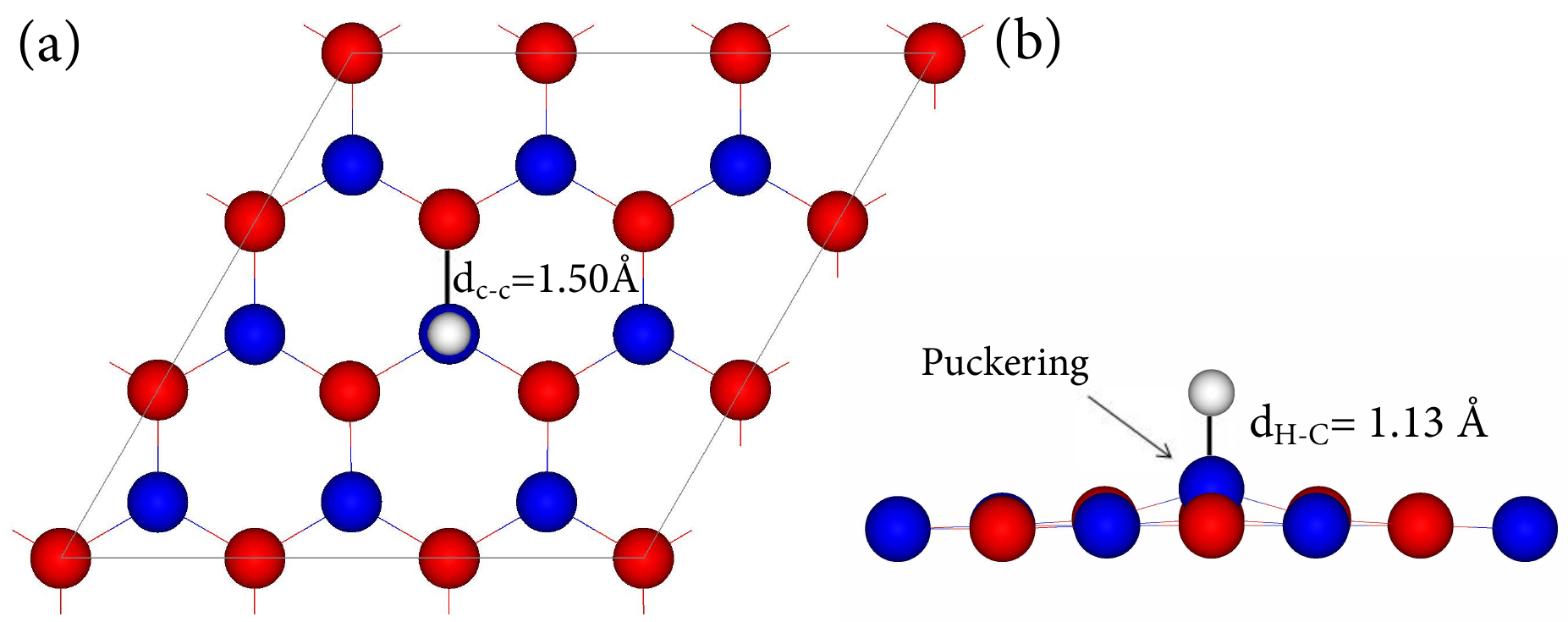}
    \else
      \includegraphics[bb = 92 86 545 742, height=3in]{H-over-graphene}
    \fi
    \caption{(Color online). 
Atomic structure of H on graphene. (a) Top and (b) side view for a $ 3\times3\times1 $ supercell.}
    \label{H-over-graphene}
  \end{center}
\end{figure}

We revisit now, for the sake of completeness, the atomic, electronic, and magnetic structure changes induced on monolayer graphene by the adsorption of a single H atom. In Fig. \ref{H-over-graphene} we present a view of the atomic structure resulting after the adsorption. This can only occur when the substrate is allowed to relax. In the stable configuration the H atom is covalently bonded to one carbon atom and is located right above this atom,
as shown in Fig. \ref{H-over-graphene}(a).
The carbon atom in the adsorption site extrudes out of the graphene plane, displaying the typical $sp^{3}$ 
hybridization to form the $\sigma$ C-H bond [see Fig. \ref{H-over-graphene}(b)]. 
For all supercell sizes we found that the bond lengths between the adsorptive carbon atom and its 
nearest neighbors increase up to 1.50{\AA}
(which is to be compared to the bond length in graphene of 1.42{\AA}).
The other bond lengths are practically unaffected and the C-H distance is always found to be 1.13{\AA}, regardless of the supercell size.
Table \ref{table:structure-H-graphene} contains a detailed account of our results compared to those found in the literature for this system.

\begin{figure}[hbp]
  \begin{center}
    \leavevmode
    \ifpdf
      \includegraphics[height=1.5in]{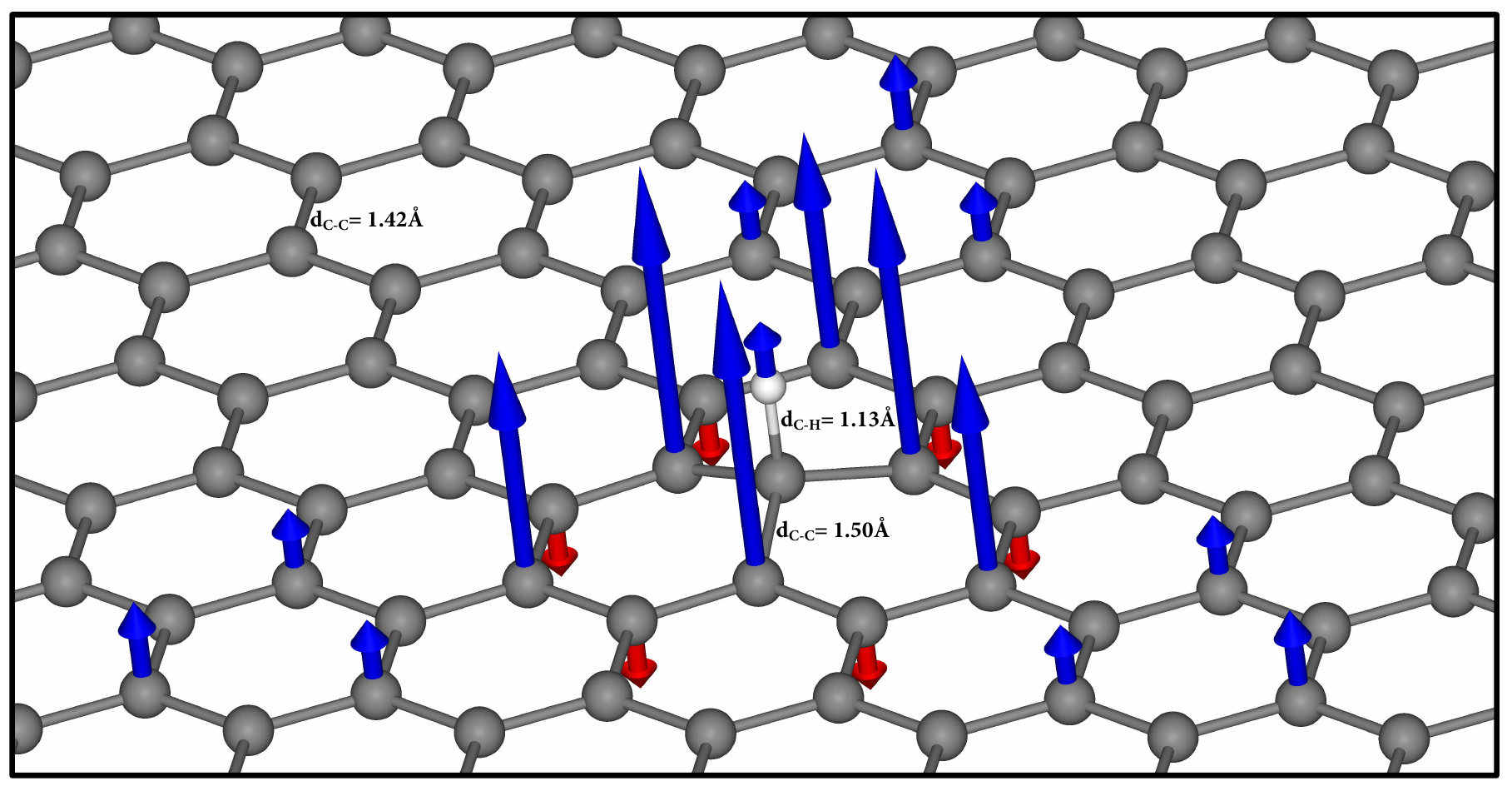}
    \else
      \includegraphics[bb = 92 86 545 742, height=3in]{Spin-Vector-pictures2}
    \fi
    \caption{(Color online). Relaxed atomic structure and spin polarization around an adsorbed H atom. Magnetic moments with opposite orientations are depicted by blue and red arrows for clarity.}
    \label{Spin-Vector-pictures}
  \end{center}
\end{figure}

The adsorption energy $ E_{a} $ for a H atom on graphene is calculated as usual
\begin{equation}
E_{a} = E_{\rm graphene+H} - ( E_{\rm graphene} + E_{\rm H} )
\end{equation}
where  $ E_{\rm graphene+H} $ denotes the total energy of the complete system and $ E_{\rm graphene} $ and $ E_{\rm H} $ 
denote the energies of the isolated graphene and H atom, respectively. We have found that
the binding energy between the H atom and a graphene monolayer increases with
increasing supercell size. A linear fit as a function of the inverse supercell size can be done for the 
calculated points which shows that the adsorption energy is about -0.98 eV in the limit of zero concentration of H atoms 
(infinite supercell size). Obviously, for a given supercell size, the binding energy of the H atom  
sublattice A is equal to the binding energy of the H atom on sublattice B $[E_{a}({\rm A}) = E_{a}({\rm B})]$.

\onecolumngrid

\begin{table}[ht]
\caption{Equilibrium height of the adsorptive carbon atom above the surface $(d_{puck})$ and adsorption energies 
$(E_a)$ for different supercell sizes and corresponding H concentration $C$. 
All the carbon atoms are allowed to relax along with the H atom.}   % title of Table
\centering
\begin{center}

\begin{tabular}{   c | c | c | c | c |  c  }                                 % centered columns (5 columns)
  %    \hline                                                                   % inserts single horizontal line
%heading

    Unit cell    & $C$ &  $d_{puck}$ (\AA{})   &  $d_{puck}$ (\AA{})     &     $E_a$ (eV)         &     $E_a $(eV)    \\  % inserts table
                 &          &        this work      &      other works         &      this work       &     other works  \\  % inserts table
  [2ex]
    \hline \hline     % inserts double horizontal lines
        $ 2\times2 $ & 0.125    & 0.359      & 0.36 \citep{Sha2002}, 0.36 \citep{Casolo2009}, 0.36 \citep{Ivanovskaya2010}&  -0.909          & -0.67 \citep{Sha2002}, -0.75    \citep{Casolo2009}, -0.83 \citep{Sljivancanin2009}, -0.85 \citep{Ivanovskaya2010}           \\
    [1ex]
    \hline
        $ 3\times3 $ & 0.056    & 0.476      & 0.41 \citep{Kerwin2008}, 0.42 \citep{Casolo2009}, 0.51 \citep{Ivanovskaya2010}        &  -0.915           & -0.76 \citep{Kerwin2008}, -0.77 \citep{Casolo2009}, -0.84   \citep{Ivanovskaya2010}      \\
    [1ex]
    \hline

    $ 4\times4 $ & 0.031        & 0.485      & 0.48   \citep{Casolo2009}, 0.49 \citep{Denis2009}, 0.58 \citep{Ivanovskaya2010}     &  -0.946           & -0.76  \citep{PhysRevLett.92.225502}, -0.85  \citep{PhysRevLett.97.186102}, -0.79  \citep{Casolo2009}, -0.89  \citep{Denis2009}, -0.89  \citep{Ivanovskaya2010}       \\               [1ex] %adds vertical space
    \hline
        $ 5\times5 $ & 0.020    & 0.500      & 0.59  \citep{Casolo2009}, 0.63  \citep{Ivanovskaya2010}     &  -0.950           & -0.82   \citep{Chen2007}, -0.84  \citep{Casolo2009},  -0.94   \citep{Ivanovskaya2010}     \\
    [1ex]
    \hline
        $ 6\times6 $ & 0.014    & 0.531      & 0.66     \citep{Ivanovskaya2010}     &  -0.956           & -0.96  \citep{Ivanovskaya2010}       \\
    [1ex]
    \hline                                                                    %inserts single line
    $\infty\times\infty$ & 0.0 & & & -0.98 & \\
    \hline
\end{tabular}

\end{center}
\label{table:structure-H-graphene}                                                        % is used to refer this table in the text
\end{table}

\twocolumngrid

\begin{figure}[!htbp]
  \begin{center}
    \leavevmode
    \ifpdf
      \includegraphics[height=2.2in]{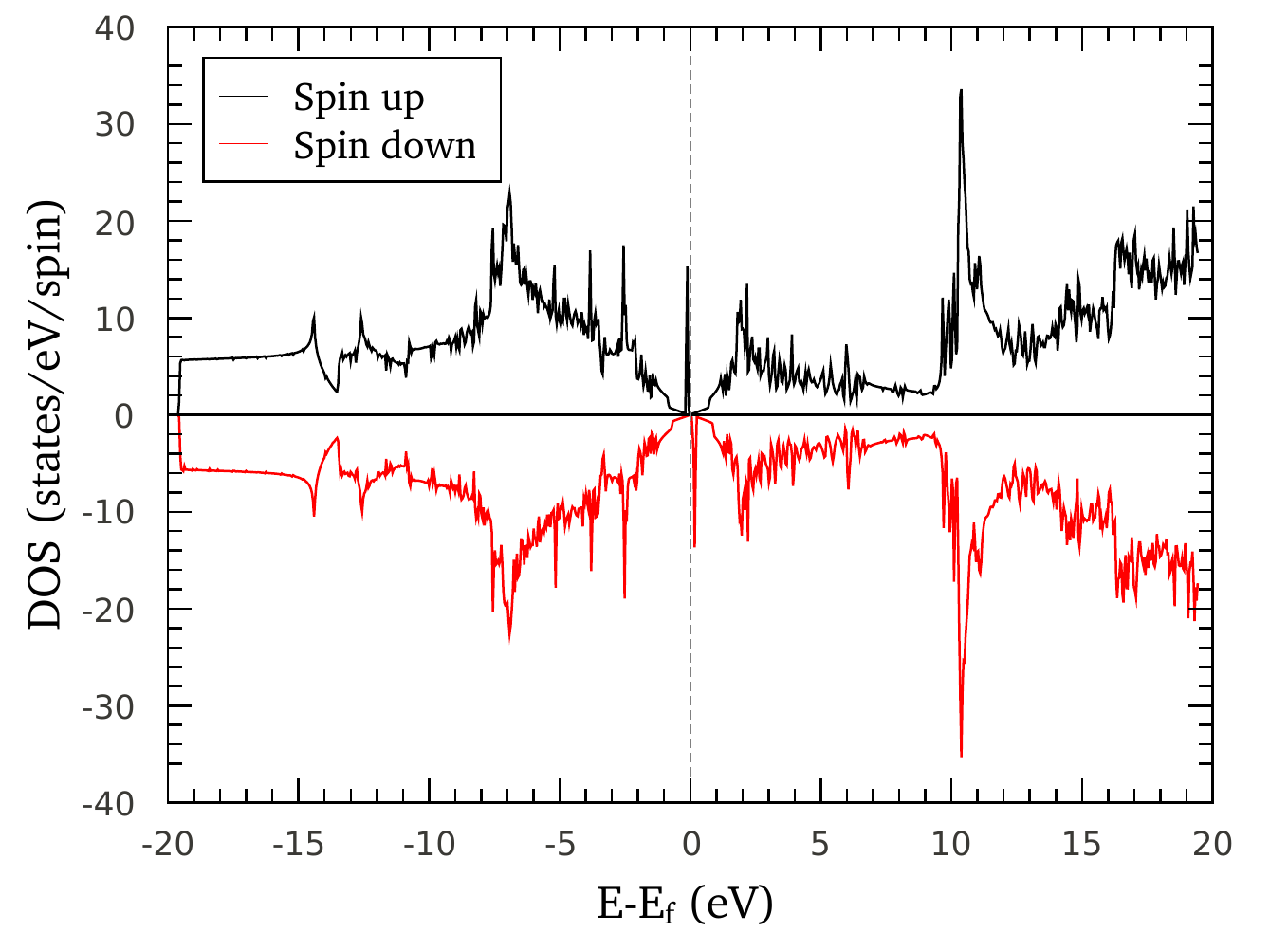}
    \else
      \includegraphics[bb = 92 86 545 742, height=3in]{DOS-H-on-graphene}
    \fi
    \caption{(Color online). Total density of states for a H atom on single-layer graphene calculated 
with a $ 6\times6\times1 $ supercell.}
    \label{DOS-H-on-graphene}
  \end{center}
\end{figure}

In agreement with previous studies we also find that the adsorption of H leads to the appearance of a staggered 
magnetization on neighbouring carbon atoms amounting to exactly 1$\mu_{\rm B}$/cell. Such spin density 
is mainly localized around the adsorptive carbon atom as shown in Fig. \ref{Spin-Vector-pictures}. 
In Fig. \ref{DOS-H-on-graphene}  we show the total density of states (DOS) for the $6\times6$ H-graphene equilibrium structure. The H adsorption causes the appearance of peak in the DOS at the Fermi level which spin-splits due to electron-electron interactions.  Remarkably, this result is compatible with Lieb's theorem for the Hubbard model on bipartite lattices \cite{PhysRevLett.62.1201}. According to such theorem, the removal of a single site in the bipartite lattice should give rise to a ground state with $S=1/2$. The covalent bond between the H atom and the C atom underneath effectively suppresses the ``site'' (the $p_z$ orbital), creating a vacancy in the underlying low-energy Hamiltonian. It is worth noticing how this result contrasts with that obtained for a vacancy. As
discussed in Ref. \onlinecite{PhysRevB.85.245443}, vacancies could in principle give rise to similar magnetic states. The difference with respect to the case of H adsorption is that vacancies tend to reconstruct and the magnetic moment generated actually vanishes for low concentrations.

%\begin{figure}[!htbp]
%  \begin{center}
%    \leavevmode
%    \ifpdf
%      \includegraphics[height=2.2in]{Formation-energy-graphene}
%    \else
%      \includegraphics[bb = 92 86 545 742, height=3in]{Formation-energy-graphene}
%    \fi
%    \caption{Adsorption energy of a H atom on graphene against different cell sizes.}
%    \label{Formation-energy-graphene}
%  \end{center}
%\end{figure}
% -----------
% -------------------------- Spin-Vector-pictures-H-on-graphene ---------------------------------------

% ------------------------------------------------------------------------

% -------------------------- DOS-H-on-graphene ---------------------------------------

\begin{figure}[!htbp]
  \begin{center}
    \leavevmode
    \ifpdf
      \includegraphics[height=2.0in]{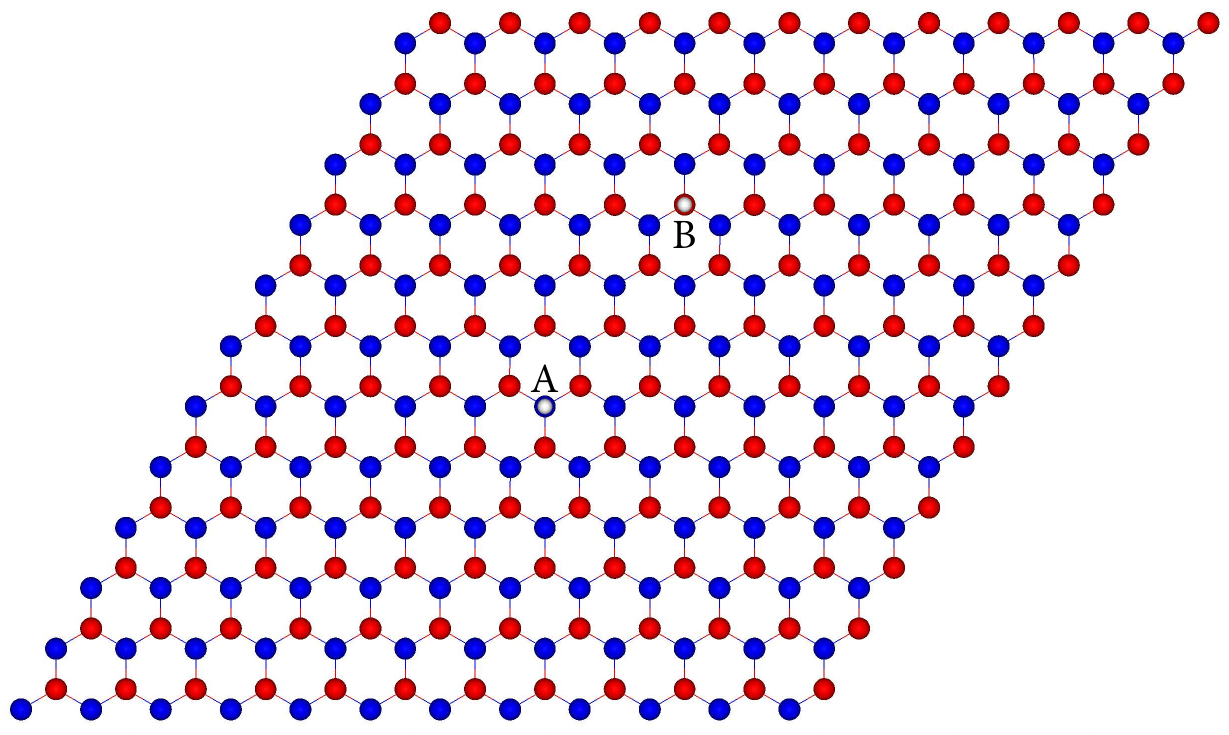}
    \else
      \includegraphics[bb = 92 86 545 742, height=3in]{Structure-paire-H-on-graphene}
    \fi
    \caption{Atomic view of a pair of H atoms on a graphene monolayer for a $ 12\times12\times1 $ supercell.}
    \label{Structure-paire-H-on-graphene}
  \end{center}
\end{figure}
% ------------------------------------------------------------------------

\subsubsection{Two hydrogen atoms}

\begin{figure}[!htbp]
  \begin{center}
    \leavevmode
    \ifpdf
      \includegraphics[height=2.7in]{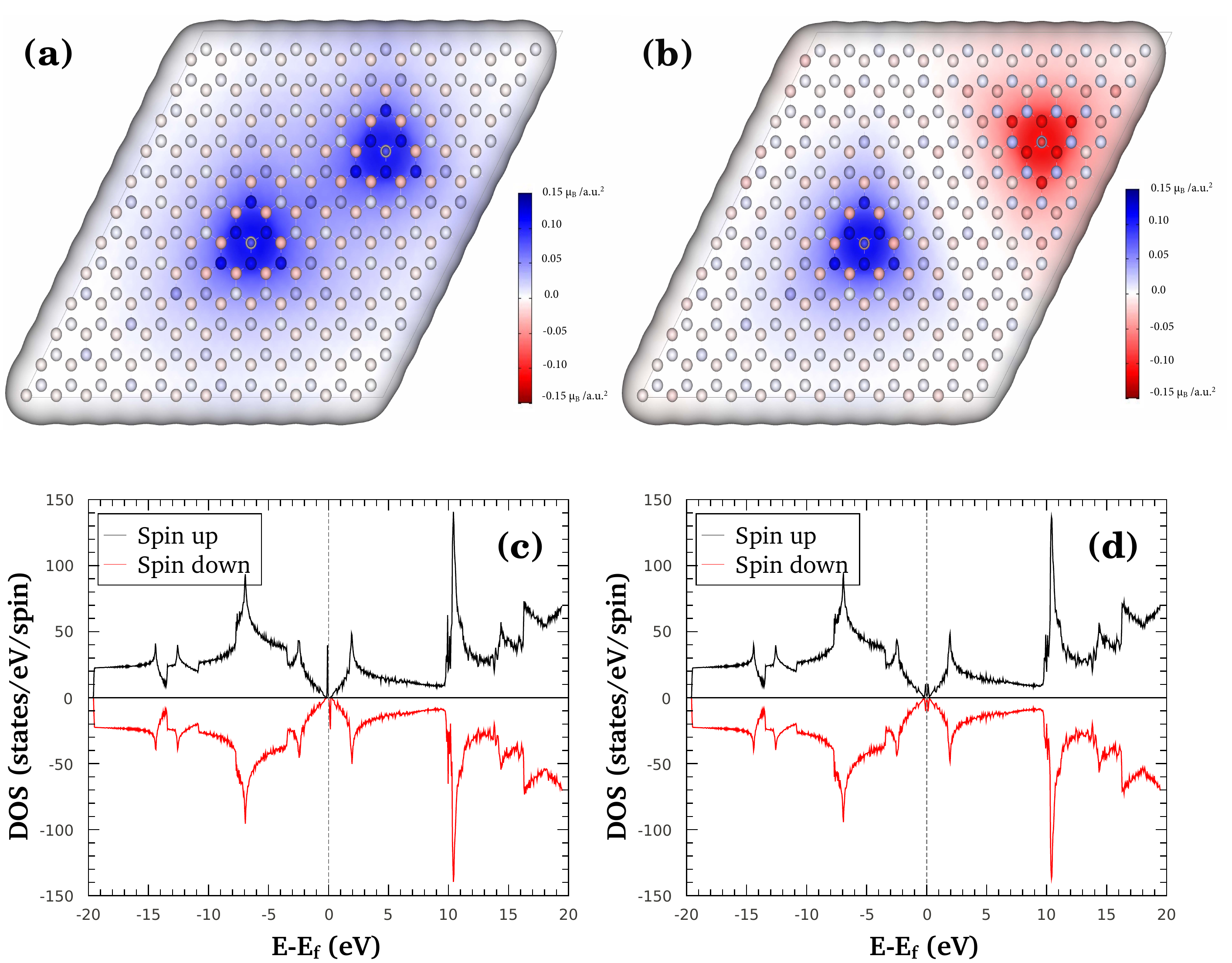}
    \else
      \includegraphics[bb = 92 86 545 742, height=3in]{DOS-AA-AB}
    \fi
    \caption{(Color online). Spin density (blue indicates up and red down) and spin-resolved total DOS for graphene monolayer with 2 H atoms sitting on AA [(a) and (c), respectively] and AB [(b) and (d), respectively] sublattices at far distances calculated with a $ 4\times4\times2 $ MP-grid for a $ 12\times12\times1 $ supercell.}
    \label{DOS-AA-AB}
  \end{center}
\end{figure}

\begin{figure}[!htbp]
  \begin{center}
    \leavevmode
    \ifpdf
      \includegraphics[height=2.5in]{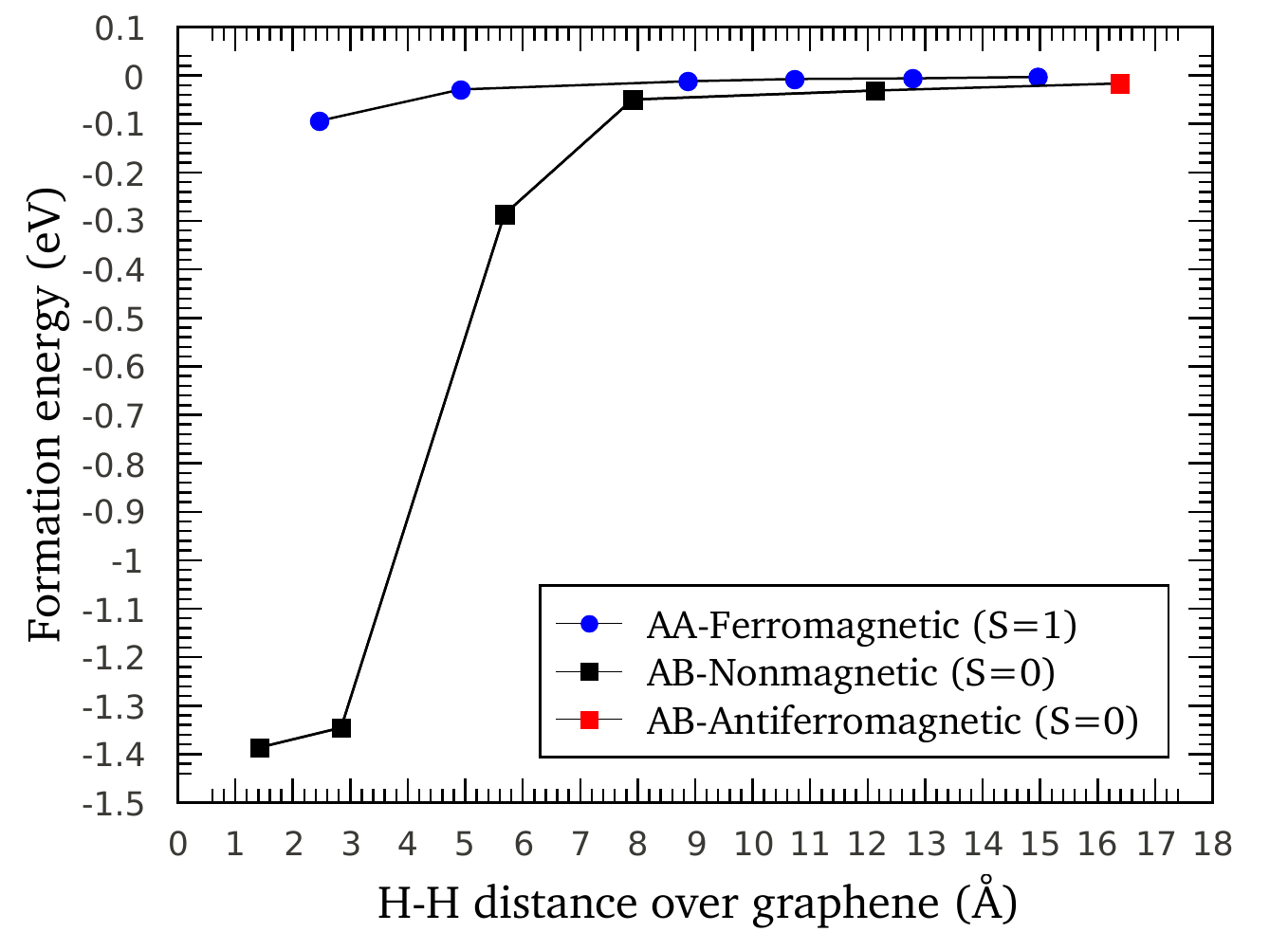}
    \else
      \includegraphics[bb = 92 86 545 742, height=3in]{2H-over-Graphene-with-142}
    \fi
    \caption{(Color online). Total energy for a pair of H atoms on a graphene monolayer ($ 12\times12\times1 $ supercell)
 relative to twice the adsorption energy of a single atom. Both AA and AB cases are shown.}
    \label{2H-on-Graphene}
  \end{center}
\end{figure}

To investigate the electronic and magnetic structure induced on graphene by two adsorbed H atoms we need
to use a $ 12\times12\times1 $ supercell. Figure \ref{Structure-paire-H-on-graphene} shows an example and illustrates
the required size of the supercell. The use of such a large supercell is essential in order 
to minimize the influence of neighboring supercells on the pair-wise properties due to the relative long-range 
interaction between the magnetic clouds induced by the H atoms. The relative extension of the magnetic clouds with respect to the supercell size is illustrated in Figs. \ref{DOS-AA-AB}(a) and (b). Test
calculations show that using larger supercells essentially gives similar results. We calculate the energetics for 
the two fundamentally different adsorption configurations. One in which the two H atoms are sitting on the 
same sublattice (AA) and the other where they are sitting on different sublattices (AB).  
The formation or adsorption energies for pairs of H atoms at various relative
distances for some AA and AB configurations are shown in
Fig. \ref{2H-on-Graphene}. (In order to see the influence of the H atoms
 on each other,  we have subtracted twice the adsorption energy of single H atom.)
We have not explored all possibilities, showing only some representative ones. Since
the magnetic cloud or localized state associated to each H atom is not isotropic, an angular dependence
is expected. The overall result remains valid though.

First, we can see that the ``interaction'' energy between atoms is always negative, i.e., the H atoms 
``attract'' each other regardless of the relative adsorption sublattices. The energy gain is the largest by placing the atoms  near each other (barely 
noticeable for the AA cases, but clearly appreciable below 1 nm for the AB ones).
This can be understood in simple terms by noticing that the H adsorption creates a localized state at 
the Fermi energy occupied by a single electron. When two states are created on different sublattices, 
these hybridize creating a bonding state that is now occupied by the two electrons 
forming a singlet state\cite{PhysRevB.77.195428}. This is essentially the reason why 
magnetic solutions only appear at long distances in the AB cases. As Fig. \ref{Structure-paire-H-on-graphene} shows, 
only for the longest possible calculated distance the H atoms retain their magnetic clouds. There the
coupling is antiferromagnetic ($S=0$), as expected from Lieb's theorem. Figures \ref{DOS-AA-AB}(b) and (d) 
show the spin-density and the spin-resolved DOS, respectively, in this case. The latter exhibits magnetic splitting
near the Fermi energy although the DOS for both spin species are identical.
On the contrary, when both atoms are on the same sublattice (AA cases) the solution is always ferromagnetic ($S=1$) 
regardless of distance [see Figs. \ref{DOS-AA-AB}(a) and (c)], but the energy gain with 
decreasing distance is very small since the localized states induced by the H atoms 
belong to the same sublattice and cannot hybridize.

% -------------------------- Energy of pair H atoms on Graphene Single Layer ---------------------------------------

% ------------------------------------------------------------------------

\vspace{50 mm} % ------------------------------------------------------------------------

\subsection{Hydrogen atoms on bilayer graphene}

\subsubsection{One hydrogen atom}

\begin{figure}[!htbp]
  \begin{center}
    \leavevmode
    \ifpdf
      \includegraphics[height=1.05in]{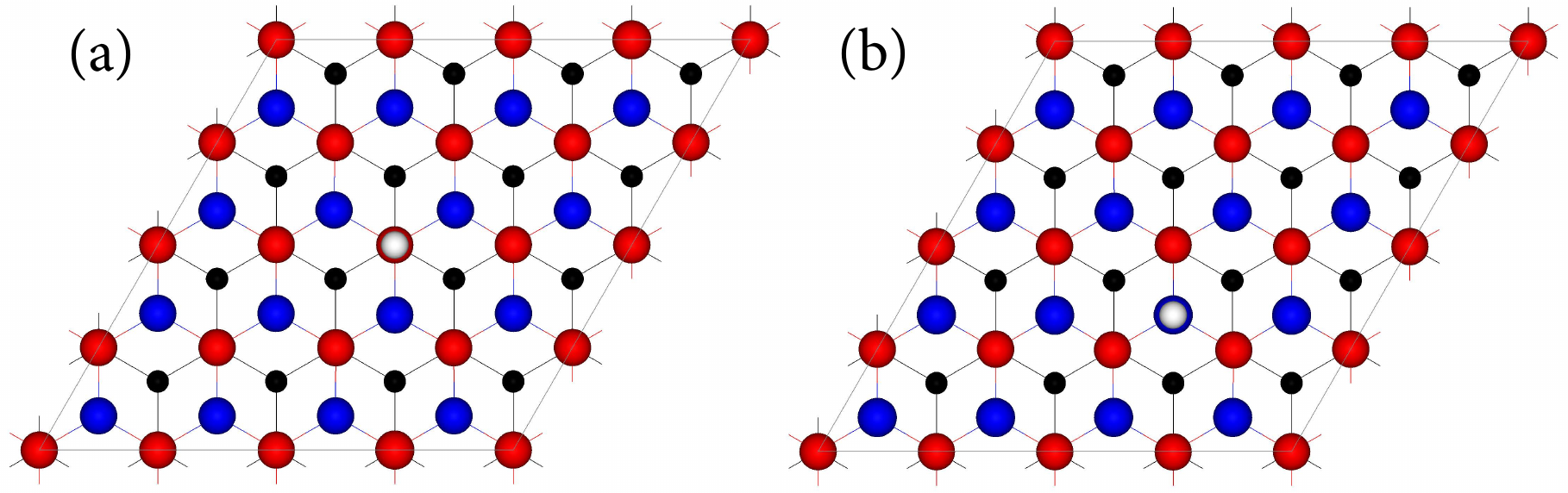}
    \else
      \includegraphics[bb = 92 86 545 742, height=3in]{structure-H-on-bilayer}
    \fi
    \caption{(Color online). Top view of the atomic structure of H on bilayer graphene for 
(a) $\alpha$ and (b) $\beta$ sites on a $ 4\times4\times1 $ supercell.}
    \label{structure-H-on-bilayer}
  \end{center}
\end{figure}

\begin{figure}[!htbp]
  \begin{center}
    \leavevmode
    \ifpdf
      \includegraphics[height=2.2in]{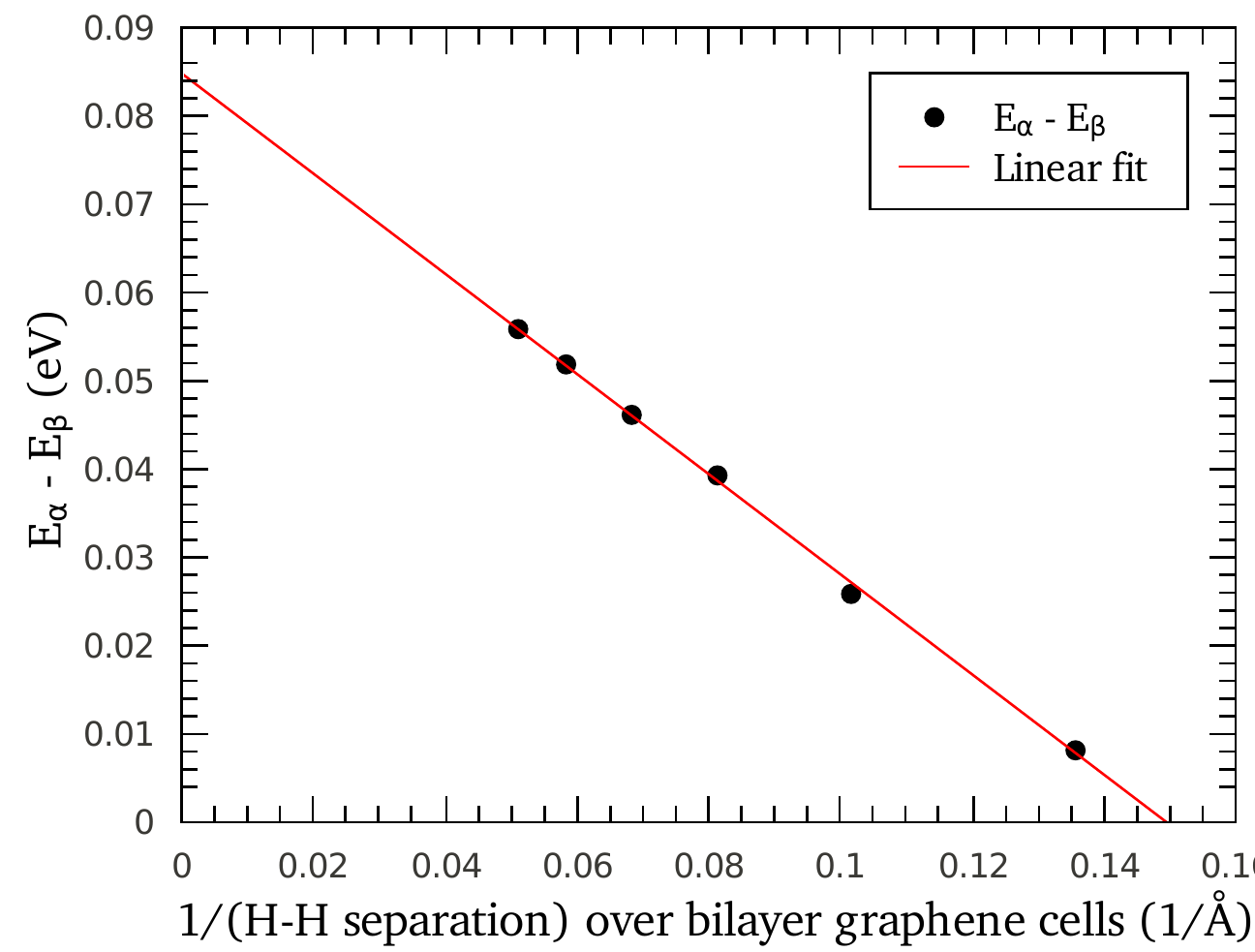}
    \else
      \includegraphics[bb = 92 86 545 742, height=3in]{difference-energy-of-H-on-bilayer}
    \fi
    \caption{(Color online). Adsorption energy difference between the two sites ($\alpha$ and $\beta$) of 
bilayer graphene against different cell sizes.}
    \label{difference-energy-of-H-on-bilayer}
  \end{center}
\end{figure}

The main focus of this work is actually to elucidate how the interactions of the graphene layers underlying the surface 
monolayer that hosts the adsorbed H atoms changes the well-established results presented in previous section. 
As we know, the most stable structure for bilayer graphene, multilayer graphene, and bulk graphite 
consists of stacked graphene monolayers following what is called Bernal stacking. 
In Fig. \ref{structure-H-on-bilayer} we present a top view of  the obtained atomic structure for the 
adsorption of a single H atom on a graphene bilayer. Here the upper layer is allowed to relax 
while the carbon atoms in the lower layer were fixed at their equilibrium position. The adsorption 
geometry of a H atom on a bilayer graphene surface is very similar to that for graphene monolayer. 
Due to the interaction between layers, however, in the bilayer graphene case (and surface graphite as shown below) 
the sublattices are not equivalent which translates into different adsorption 
energies $[E_{a}(\alpha) > E_{a}(\beta)]$. (In order to make clear that the surface sublattices 
are not equivalent anymore, we change the labels A and B to $\alpha$ and $\beta$ from now on.) 
In Fig. \ref{difference-energy-of-H-on-bilayer} we show the H adsorption energy difference 
between $\alpha$ and $\beta$ sites [$\Delta E=E_{a}(\alpha)-E_{a}(\beta)$] for different supercell sizes of 
the graphene bilayer. $\Delta E$ increases linearly with the supercell size, 
extrapolating to $\approx 85$ meV for infinitely large supercells. Importantly, 
the induced magnetic moment is not affected by the presence of the second graphene layer.

% -------------------------- Structure ---------------------------------------

% ------------------------------------------------------------------------

% -------------------------- difference-energy-of-H-on-bilayer ---------------------------------------

% ------------------------------------------------------------------------

% -------------------------- DOS-H-on-Bilayer ---------------------------------------

%\begin{figure}[!htbp]
 % \begin{center}
 %   \leavevmode
 %   \ifpdf
 %     \includegraphics[height=2.7in]{DOS-Bilayers}
 %   \else
 %     \includegraphics[bb = 92 86 545 742, height=3in]{DOS-Bilayers}
 %   \fi
 %   \caption{Total density of states calculated for a) 4x4x1 supercell b)5x5x1 supercell c) 6x6x1 supercell d) 7x7x1 supercell.}
 %   \label{DOS-Bilayers}
 % \end{center}
%\end{figure}

% -------------------------------- End DOS --------------------------------------

\subsubsection{Two hydrogen atoms} % --------------------------------------------------------------------------------------------------------------

% -------------------------- difference-energy-of-H-on-bilayer ---------------------------------------

\begin{figure}[!htbp]
  \begin{center}
    \leavevmode
    \ifpdf
      \includegraphics[height=2.2in]{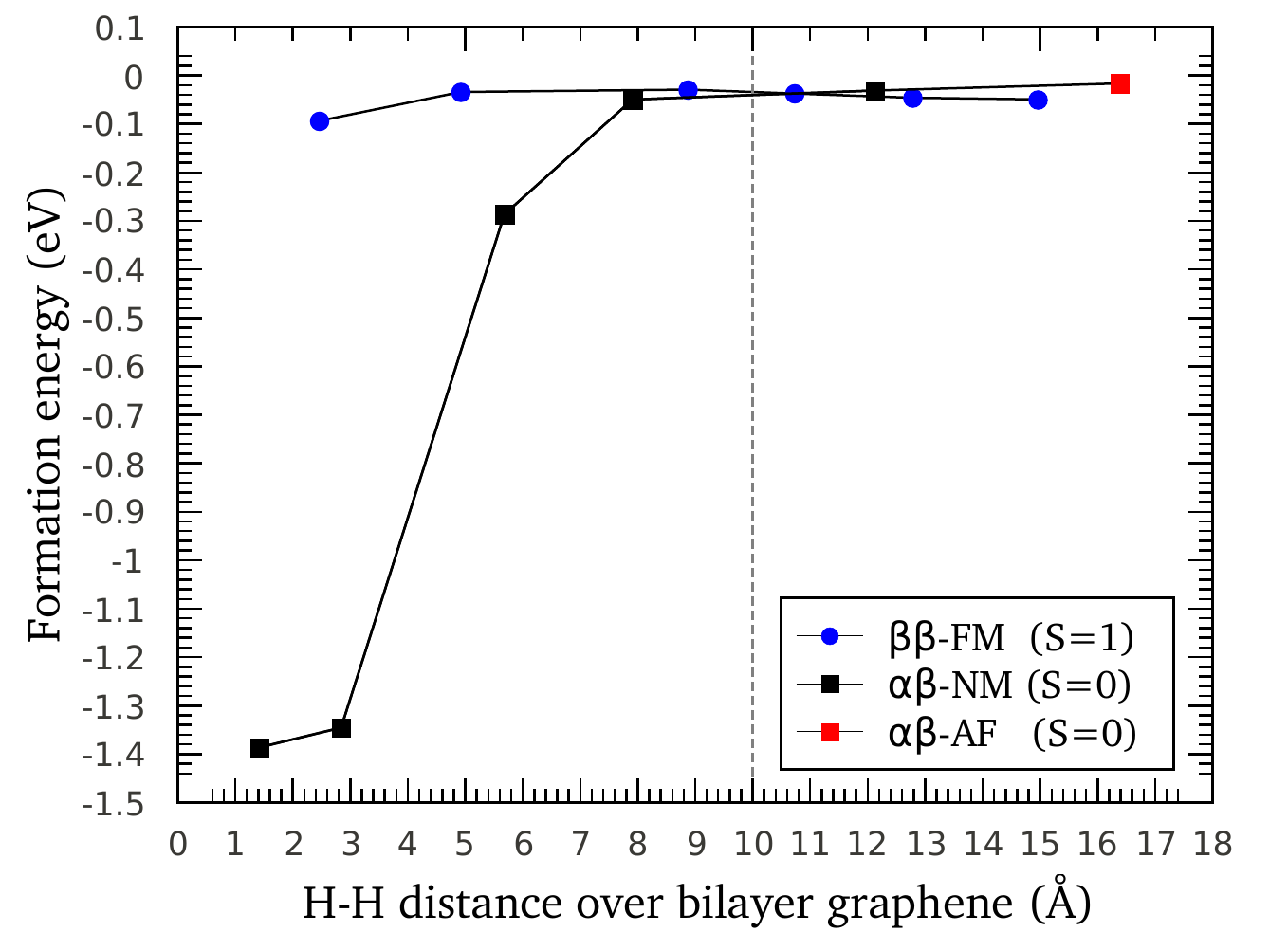}
    \else
      \includegraphics[bb = 92 86 545 742, height=3in]{2H-over-Bilayer-with-142}
    \fi
    \caption{(Color online). Total energy for two H atoms on bilayer graphene as a function of distance.}
    \label{2H-over-Bilayer}
  \end{center}
\end{figure}
% ------------------------------------------------------------------------

As shown in previous sections, to properly investigate the interaction between two adsorbed H atoms, 
one requires very large supercells. A similar study in the bilayer case is computationally prohibited. 
Here we adopt a different approach. We assume that the attractive interaction between H atoms is not affected 
by the underlying graphene layer. This is not a strong assumption since the interaction between 
layers is mainly of van der Waals type while the origin of the magnetic structure changes induced on graphene 
by the adsorbed H are of kinetic and exchange type. We now simply shift the AA pair energy  
shown in Fig. \ref{2H-on-Graphene} by the energy difference between $\alpha$ and $\beta$ adsorption sites, $\Delta E$. 
 There are two possibilities here. One is to use the value of $\Delta E$ obtained in the limit of 
infinitely large supercells. The other is to use a value of $\Delta E$ that changes with the 
distance between H atoms. This can be estimated from the calculation for a given supercell size that 
approximately corresponds to such distance.  Either choice obviously favors adsorption on the same 
sublattice ($\beta$ in this case) when the H atoms are sufficiently far apart and the intra-layer 
interactions are weakened. There are not significant differences between both choices and the result  
for the second one is shown in Fig. \ref{2H-over-Bilayer}.  As can be seen, the pairs of H atoms prefer to sit on the same 
sublattice for distances longer that $\approx 1$ nm, thus favouring a ferromagnetic state on the surface layer 
for a maximum coverage $C$ of around 0.05.  

We note that, although the thermodynamically
most stable situation is when H atoms approach one another forming pairs or clusters, the attraction between H atoms may be 
counteracted by the diffusion barriers, particularly at low temperatures\cite{Sljivancanin2009}. 
Understanding the dynamics resulting from diffusion processes (and desorption ones for that matter)
is of great importance to determine actual hydrogenation patterns, but this lies
beyond the scope of this work.  Kinetic Monte Carlo studies have been recently carried out\cite{Mohammed-thesis}, 
indicating that, since desorption rates turn out to be smaller than diffusion ones, 
metastable states where all H atoms stay, at least temporarily, adsorbed on the same sublattice are possible.

% ------------------------------------------------------------------------
% --------------- Hydrogen atoms on multilayer graphene ------------------
% ------------------------------------------------------------------------

\subsection{Hydrogen atoms on the surface of graphite}

\begin{figure}[!htbp]
  \begin{center}
    \leavevmode
    \ifpdf
      \includegraphics[height=1.5in]{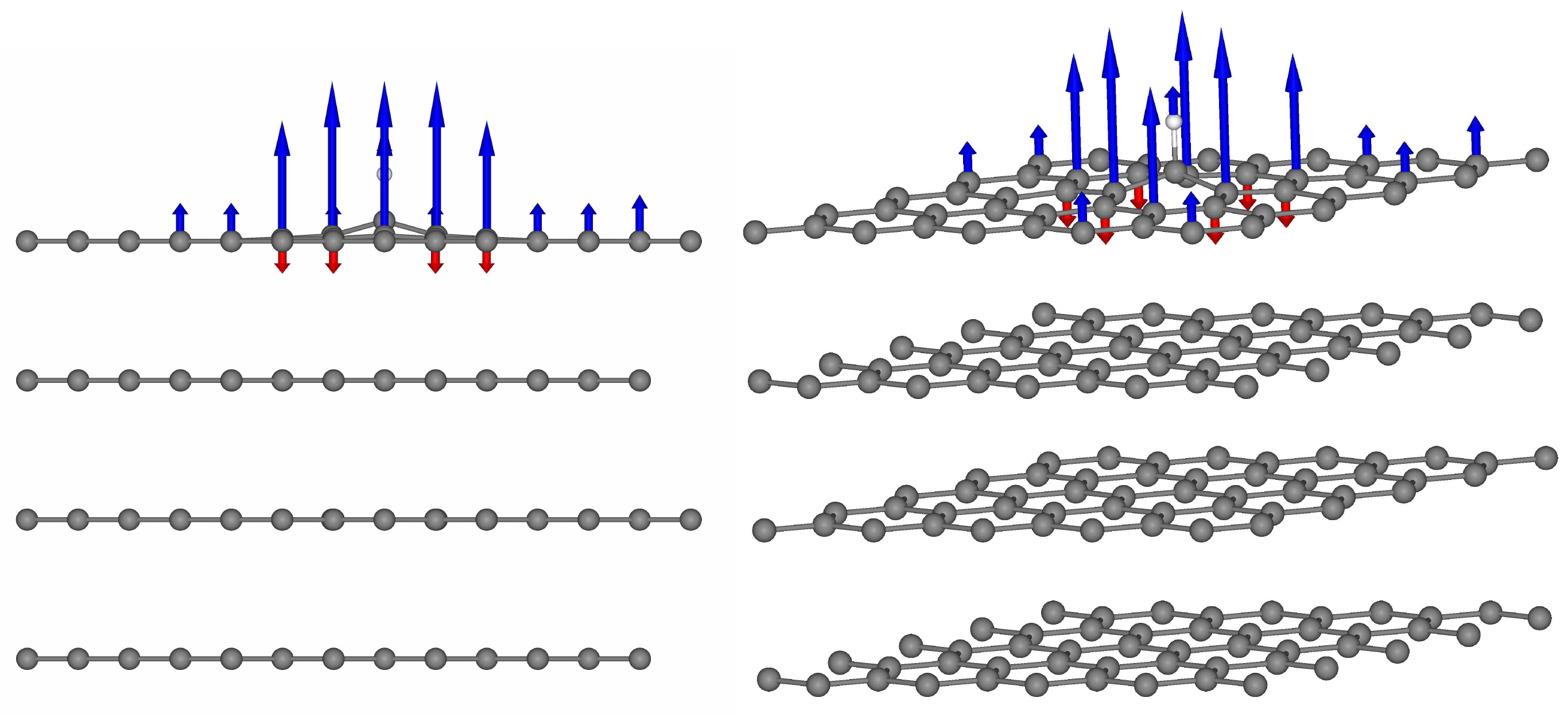}
    \else
      \includegraphics[bb = 92 86 545 742, height=1.5in]{T2-5x5-4-layers}
    \fi
    \caption{(Color online). Relaxed atomic structure and spin polarization around an adsorbed H atom at 
$\beta$ site on a 4-layer graphene surface. Magnetic moments are depicted by blue(red) arrows for 
spin-up(spin-down) for clarity.}
    \label{T2-5x5-4-layers}
  \end{center}
\end{figure}

We have mentioned in passing that the magnetic moment induced in a single graphene monolayer
by the H adsorption survives when a second layer is added to form a bilayer. This is result is not
necessarily obvious, neither is the fact that H adsorbed on a graphite surface may induce a magnetic moment as well. 
As discussed in Ref. \cite{PhysRevB.85.245443}, vacancies tend to loose the magnetic moment because the 
electron-hole symmetry is severely broken and the localized state hosting the unpaired electron is not 
exactly placed at the Fermi energy. A similar effect could take place here. To discard this possibility we
have evaluated the atomic and magnetic structures of a H atom adsorbed on graphite (represented by up to a 
five-layer graphene structure). In Fig. \ref{T2-5x5-4-layers} we present the atomic structure 
determined after the adsorption of a H atom on the surface. Here, also, the upper layer is allowed to relax 
while the carbon atoms in the underlying layers were fixed at their equilibrium position. The adsorption of the 
H atom leads to the formation of a spin density on neighboring carbon atoms, again amounting to exactly  
1$\mu_{\rm B}$/cell. Such spin density  is mainly localized on the adsorptive layer, as shown in 
Fig. \ref{T2-5x5-4-layers}. Due to the stacking order in the multi-layer graphene structure, 
the sublattices are, again, inequivalent $[E_{a}(\alpha) > E_{a}(\beta)]$ for adsorption.
In Table (\ref{table:H-on-multilayer}) we show the adsorption energy difference  $\Delta E$  for 
a $5\times5$ supercell size against different numbers of graphene layers. This converges very quickly 
with the number of layers so that the results obtained in previous section remain valid here: H atoms 
adsorbed on a graphite surface prefer to locate themselves on the same sublattice when sufficiently far apart 
from each other and induce a ferromagnetic state on the surface. The Curie temperature of this novel 
ferromagnet is analysed in the following section.

% -------------------------- Hydrogen on multilayer Table --------------------------------------
\begin{table}[ht]
\caption{Energy difference $(\Delta E)$ between $\alpha$ and $\beta$ adsorption sites for a $5\times5$ supercell size against different numbers of graphene layers.}   % title of Table
\centering
\begin{center}

\begin{tabular}{   c | c   }                                 % centered columns (3 columns)
  %    \hline                                                                   % inserts single horizontal line
%heading

    No. of layers   &     $\Delta E=E_a({\alpha})-E_a({\beta}$) (eV)    \\  % inserts table
  [2ex]
    \hline \hline     % inserts double horizontal lines
          $ 1 $ &  0.00000           \\
    [1ex]
    \hline
          $ 2 $ &  0.03930            \\
    [1ex]
    \hline
          $ 3 $ &  0.03798            \\
    [1ex] %adds vertical space
    \hline
          $ 4 $ &  0.03866            \\
    [1ex]
    \hline
          $ 5 $ &  0.03857            \\
    [1ex]
    \hline
          $ 6 $ &  0.03871            \\
    [1ex]
    \hline                                                                  %inserts single line
\end{tabular}

\end{center}
\label{table:H-on-multilayer}                                                        % is used to refer this table in the text
\end{table}
% ------------------------------------------------------------------------

% -------------------------- T2-5x5-4-layers ---------------------------------------

% ------------------------------------------------------------------------

%In Figure (\ref{DOS-H-on-multilayer}),  we show the total density of state (DOS) of the $4\times4$ supercell graphene 5-layers equilibrium structure. It is obtain from the figure that the hydrogen adsorption causes the appearance of a double peak in the DOS, symmetrically placed around the Fermi level.

% -------------------------- DOS-H-on-multilayer-4x4-5-layers ---------------------------------------

%\begin{figure}[!htbp]
 % \begin{center}
  %  \leavevmode
   % \ifpdf
    %  \includegraphics[height=2.2in]{DOS-H-on-multilayer}
   % \else
    %  \includegraphics[bb = 92 86 545 742, height=2.2in]{DOS-H-on-multilayer}
%    \fi
 %   \caption{Total density of states for hydrogen atom on five-layers graphene calculated with a $ 60\times60\times2 $ MP-grid for a $ 4\times4\times1 $ supercell.}
  %  \label{DOS-H-on-multilayer}
  %\end{center}
%\end{figure}
% ------------------------------------------------------------------------

% ------------------------------------------------------------------------
% ------------------- Curie Temperature Calculation ----------------------
% ------------------------------------------------------------------------

\section{Curie Temperature}

\begin{figure}[!htbp]
  \begin{center}
    \leavevmode
    \ifpdf
      \includegraphics[height=2.2in]{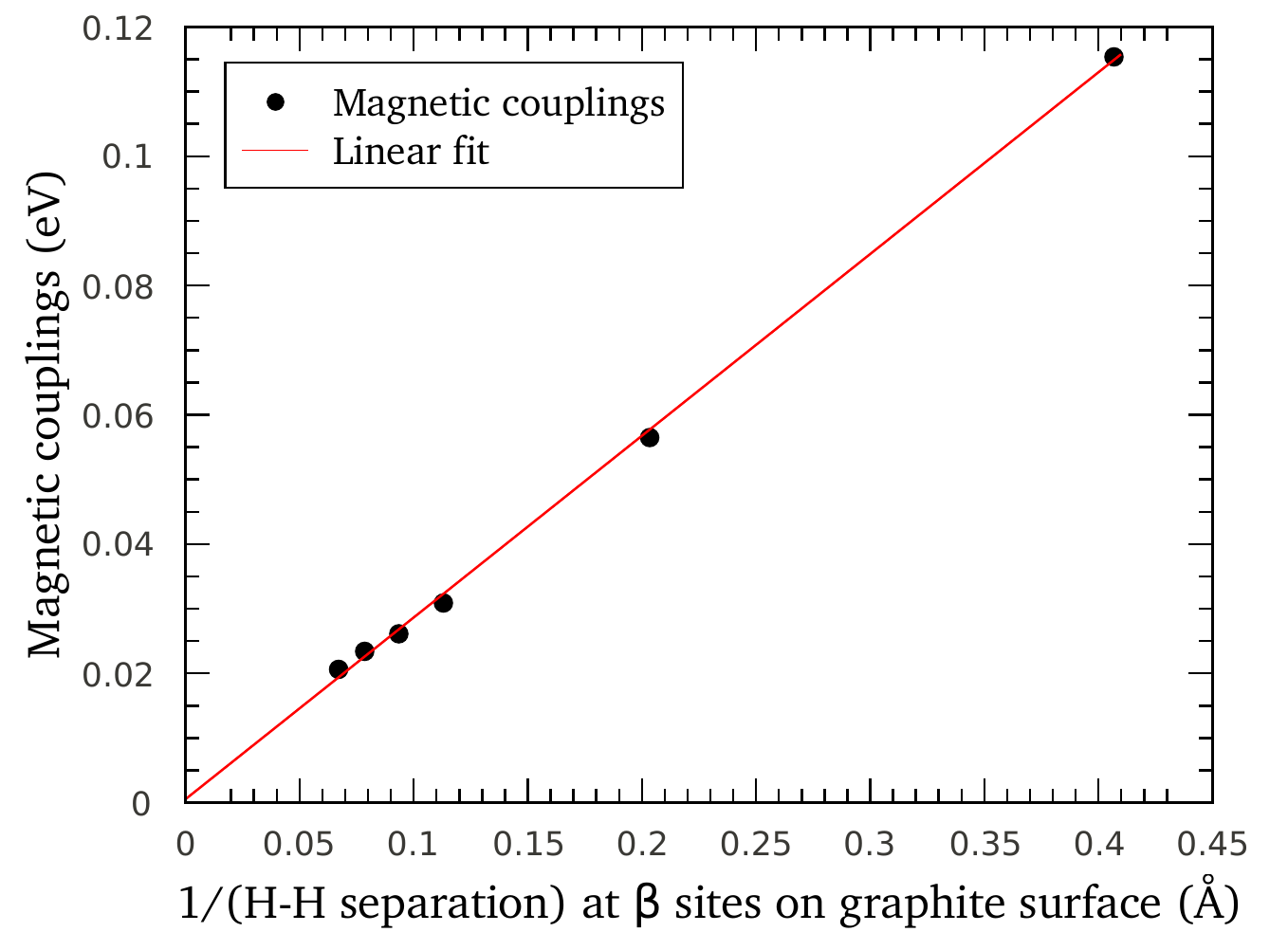}
    \else
      \includegraphics[bb = 92 86 545 742, height=3in]{Fit-Exchange-Energy}
    \fi
    \caption{(Color online). Exchange energy for a pair of H atoms adsorbed on the same sublattice .}
    \label{Fit-Exchange-Energy}
  \end{center}
\end{figure}

Our results show that the adsorption of H atoms on a graphite surface may induce, at low concentrations, 
ferromagnetically coupled  spin densities distributed around the adsorbed H atoms. In the diluted regime, 
the extension of the polarization cloud may be considered small compared to the mean distance between H atoms; 
therefore, to study the collective magnetic properties of the system  we will use the following Ising-like  model
Hamiltonian:
\begin{equation}
H = -\frac{1}{2}\sum_{ij} J_{ij} p_iS_i p_j S_j,
\label{SpinHamiltonian}
\end{equation}
where $S_i$ and $S_j$ are two discrete spin variables ($\pm 1$) at sites $i$ and $j$ of a given sublattice (say $\beta$)
of the graphite surface.
The random variables $p_i$ and $p_j$ represent the occupation of one carbon atom with a H atom. 
These can take the values 1 (occupied) or 0 (unoccupied). These discrete random variables are drawn 
from a probability density function:
\begin{equation}
    \rho(p)=(1-c)\delta(p)+c\delta(p-1),
\label{probability_density}
\end{equation}
where $c$ in $[0,1]$ is related to the graphene lattice coverage by $C=c/2$. The maximum 
coverage in our case is thus $C=0.5$ although, as explained above, it is only meaningful for $C<0.05$. 
The adimensional concentration parameter 
$c$ defines a mean distance between H atoms $\ell=\frac{1}{\sqrt{c}}$ in units of the lattice parameter $a$. 
$J_{ij}$ is the magnetic coupling constant between two magnetic moments at sites $i$ and $j$.  
The coupling constant is defined as the total energy difference between the antiparallel (AFM) and parallel (FM) 
alignment of an AA pair:
\begin{equation}
J_{ij} = (E^{\rm FM})_{ij}-(E^{\rm AFM})_{ij}.
\end{equation}
In Fig. \ref{Fit-Exchange-Energy} we show he magnetic coupling $J_{ij}$ as obtained from our DFT calculations
in the configurations 
shown in Fig. \ref{2H-on-Graphene}. The exchange energy presents a slow linear decrease  with the inverse of the  
H-H pair separation $J_{ij}=\frac{J_0 a}{r_{ij}}$  where $J_0=0.0576$ eV\cite{Dani}, and $r_{ij}$ is the distance between H atoms at sites $i$ and $j$. As expected, it extrapolates to 0 eV in the infinite separation limit.

% -------------------------- magnetic couplings Jex ---------------------------------------

% ------------------------------------------------------------------------

% -------------------------- ABS-Magnetization-vs-Temperature ---------------------------------------

\begin{figure}[!htbp]
  \begin{center}
    \leavevmode
    \ifpdf
      \includegraphics[height=1.3in]{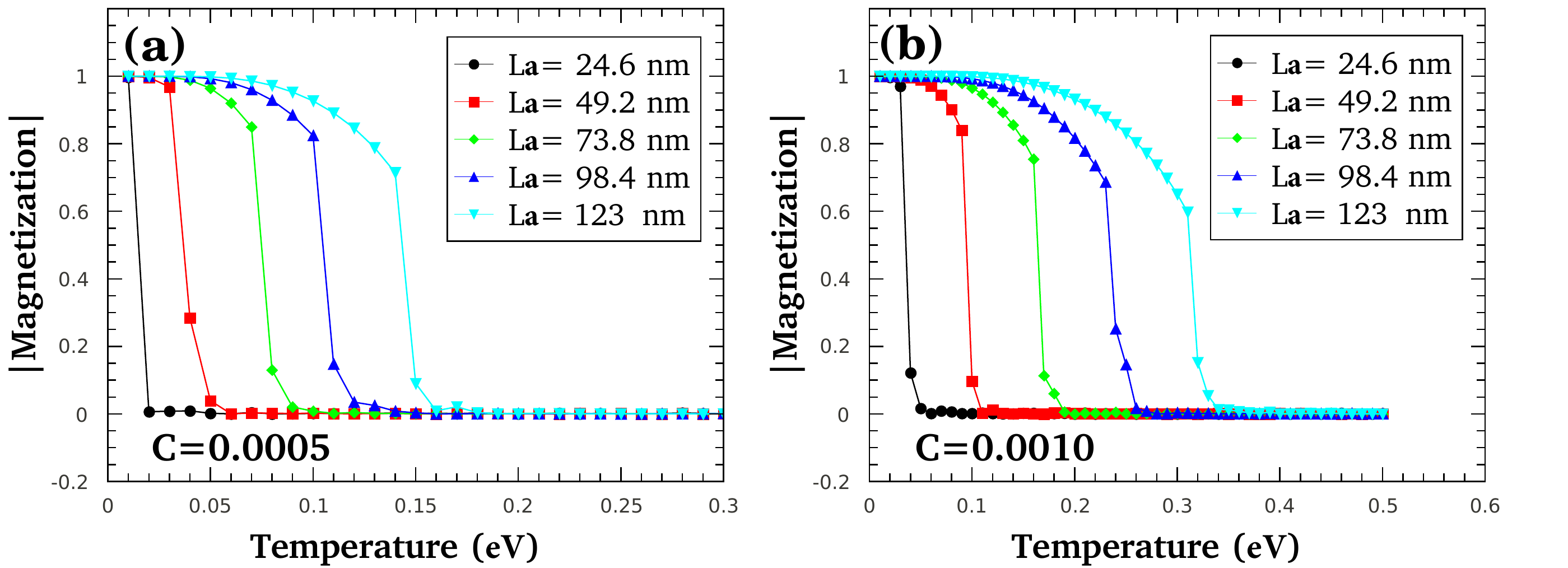}
    \else
      \includegraphics[bb = 92 86 545 742, height=2.6in]{ABS-Magnetization-vs-Temperature}
    \fi
    \caption{(Color online). Absolute magnetization per spin for supercells sizes in the range
$L= 24.6$ nm ($L= 100$ supercell units) and 123 nm ($L= 500$ supercell units),
using concentrations (a) $C= 0.0005$ and (b) 0.0010.}
    \label{ABS-Magnetization-vs-Temperature}
  \end{center}
\end{figure}
% ------------------------------------------------------------------------

To study the magnetic ordering in this system we have used a Monte Carlo (MC) algorithm\citep{Matsumoto02022001}. 
   We have simulated very diluted triangular lattices with $L\times L$ cells with $L$ in the range $L=80-1280$. 
Considering that a $1/r$ coupling has always longer range than the size of the system,
we have decided to apply open boundary conditions.
To make contact with realistic experimental realizations\cite{brihuega}, we have performed 
simulations at very low concentrations $C=0.0005, 0.0006, 0.0007, 0.0008 , 0.0009 $ and $0.0010$  
($C=1$ means full coverage of the graphite surface with H atoms). Note also that our simulations are performed 
in the range $ \frac{L}{\ell} \gg 1$. The thermal averaging took 50000 MC measurements, after allowing 1000 steps for thermalization. Average over 50 random realizations of the H distribution was taken.

In Fig. \ref{ABS-Magnetization-vs-Temperature} we show the thermal average of the magnetization absolute value  $|M|$ 
for two concentrations ($C= 0.0005$ and $C= 0.0010$) and cell sizes of  $L= 24.6$ nm ($L= 100$ supercell units), 
49.2 nm, 73.8 nm, 98.4 nm, and 123 nm ($L= 500 $ supercell units).  
The abrupt supression of $|M|$ signals the approximate value of the ordering or Curie temperature $T_{\rm C}$. 
However, this ordering temperature seems to increase with the system size. In the thermodynamic limit 
this behavior extrapolates to an infinite value (i.e., a finite  magnetization at any finite temperature).

We discuss now that this is an  intrinsic property of the system, consequence of the long-range coupling 
between the induced magnetic moments. To study this we compute the Binder cumulant, used conventionally 
for an accurate determination of the critical temperature in MC simulations of statistical systems. 
The Binder cumulant is the fourth order cumulant of the order parameter 
distribution \citep{1981ZPhyB..43..119B, PhysRevLett.47.693}, which is defined as
\begin{equation}
U_{L}(T) = \frac{1}{2}\biggl[3 - \frac{\langle \bar M^{4}\rangle}{\langle \bar M^{2}\rangle^{2}}\biggl],
\label{binder}
\end{equation}
where $\langle \bar M^{2}\rangle$ and $\langle \bar M^{4}\rangle$ are the second and fourth moments of the magnetization distribution, with the brackets $\langle...\rangle$ and the bar denoting thermal and sample averaging.

The finite-size scaling argument states that, close to a critical point, a thermal average of a generic quantity scales as 
\begin{equation}
\langle O \rangle =L^{\mu} g_O(L/\xi),
\end{equation}
where $ L$ is the system size, $\mu$ a critical exponent, and $\xi$ is the temperature dependent correlation 
length which can be considered adimensional or in units of $a$. Close to the critical point, 
it scales as $\xi(T) \sim (T-T_{\rm C})^{-\nu}$ . It is well known that several physical properties have 
important finite size corrections which makes the determination of $T_{\rm C}$ difficult. 
However, if we specifically consider the scaling of the moments of the order parameter:
\begin{equation}
  \langle M^{2n} \rangle=L^{{2 n \beta}{\nu}} g_{M^{2n}}(L/\xi)
\end{equation}
and substitute in the Binder parameter expression of Eq. \ref{binder} we get $U_{L}(T)=U(L/\xi(T))$, which 
is size independent at the critical point. At large temperatures the histogram of the magnetization 
is expected to be a Gaussian distribution and therefore $U_{L}(T \rightarrow \infty)=0$. On the other hand, 
in the zero temperature limit, the magnetization distribution function reduces to two delta peaks 
at opposite values of the saturation magnetization
and hence  $U_{L}(T \rightarrow 0)=1$. If a system has a well-defined second order phase transition
at a finite temperature,  the finite-size analysis of the Binder parameter $U_{L}(T)$  will show a family
 of decreasing functions of the temperature, all of them crossing, to a very good approximation, at $T_{\rm C}$.

% -------------------------- Binder-cumulant-vs-T ---------------------------------------
\begin{figure}[!htbp]
  \begin{center}
    \leavevmode
    \ifpdf
      \includegraphics[height=1.30in]{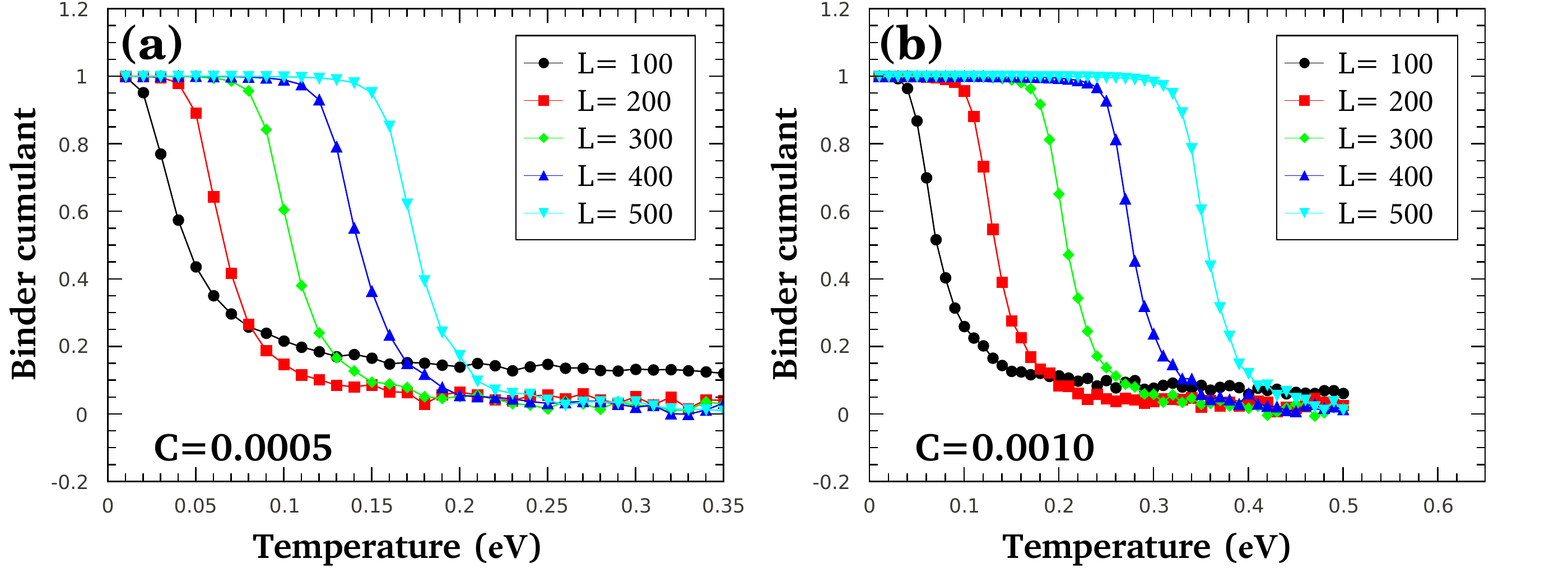}
    \else
      \includegraphics[bb = 92 86 545 742, height=2.6in]{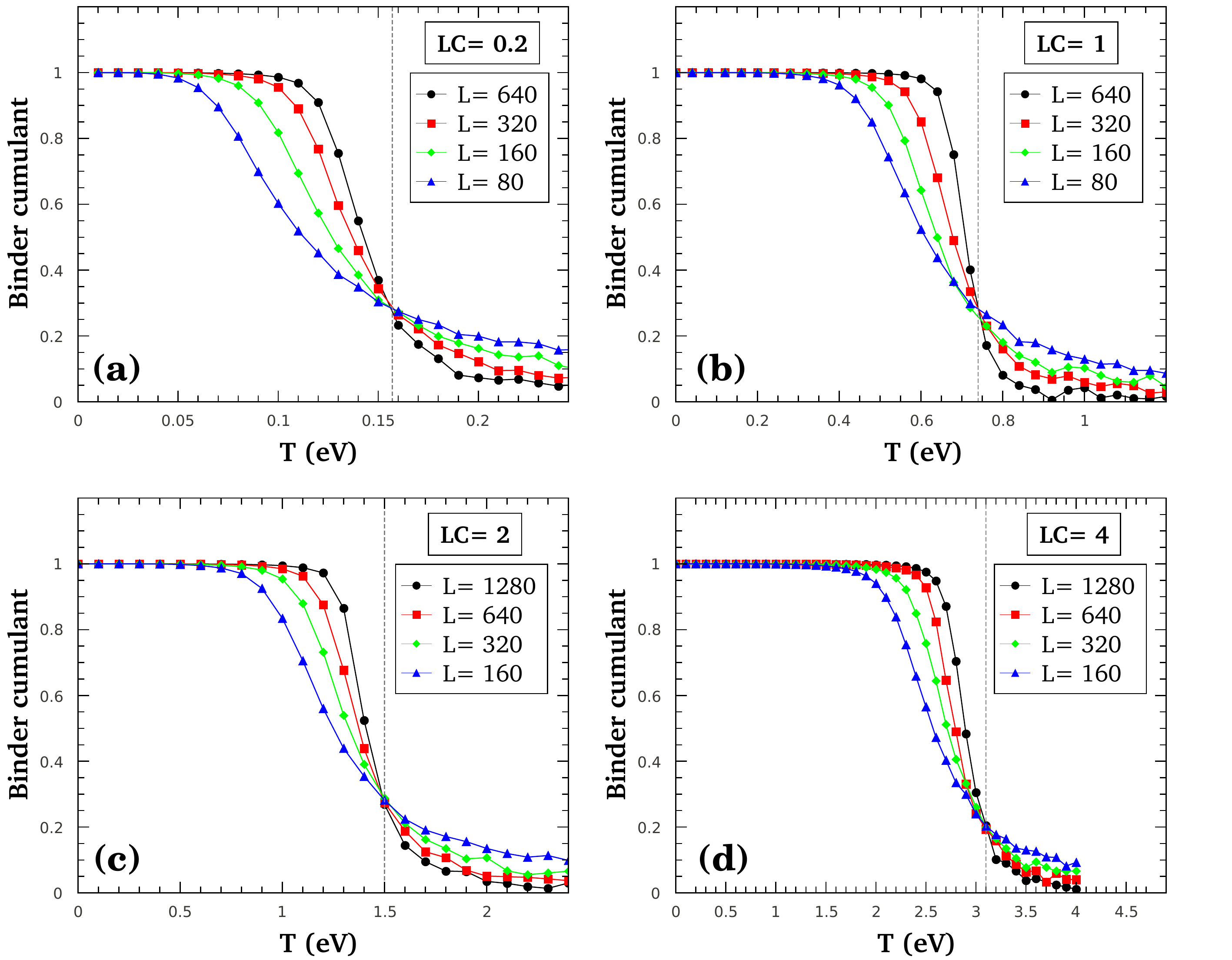}
    \fi
    \caption{(Color online). Fourth-order cumulant for supercells sizes in the range 
$L= 24.6$ nm ($L= 100$ supercell units), and 123 nm ($L= 500$ supercell units), 
using concentrations (a)$ C= 0.0005$, and (b) 0.0010}
    \label{Binder-cumulant-vs-T-for-C}
  \end{center}
\end{figure}
% ------------------------------------------------------------------------

\begin{figure}[!htbp]
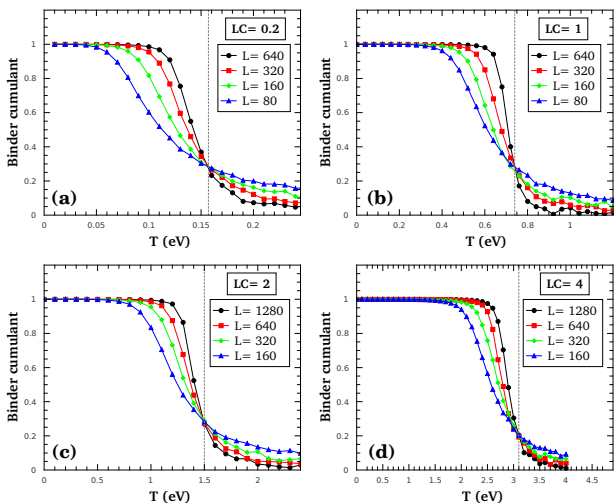

  \begin{center}
    \leavevmode
    \ifpdf
      \includegraphics[height=2.60in]{Binder-cumulant-vs-T-for-LC}
    \else
      \includegraphics[bb = 92 86 545 742, height=2.6in]{Binder-cumulant-vs-T-for-LC}
    \fi
    \caption{(Color online). Fourth-order cumulant for (a) $LC=$ 0.2, (b)$ LC=$ 1.0, (c)$ LC=$ 2.0, and (d)$ LC=$ 4,
 using supercell sizes$ L=$ 80, 160, 320, 460, and 1280.}
    \label{Binder-cumulant-vs-T-for-LC}
  \end{center}
\end{figure}

\begin{figure}[!htbp]
  \begin{center}
    \leavevmode
    \ifpdf
      \includegraphics[height=2.2in]{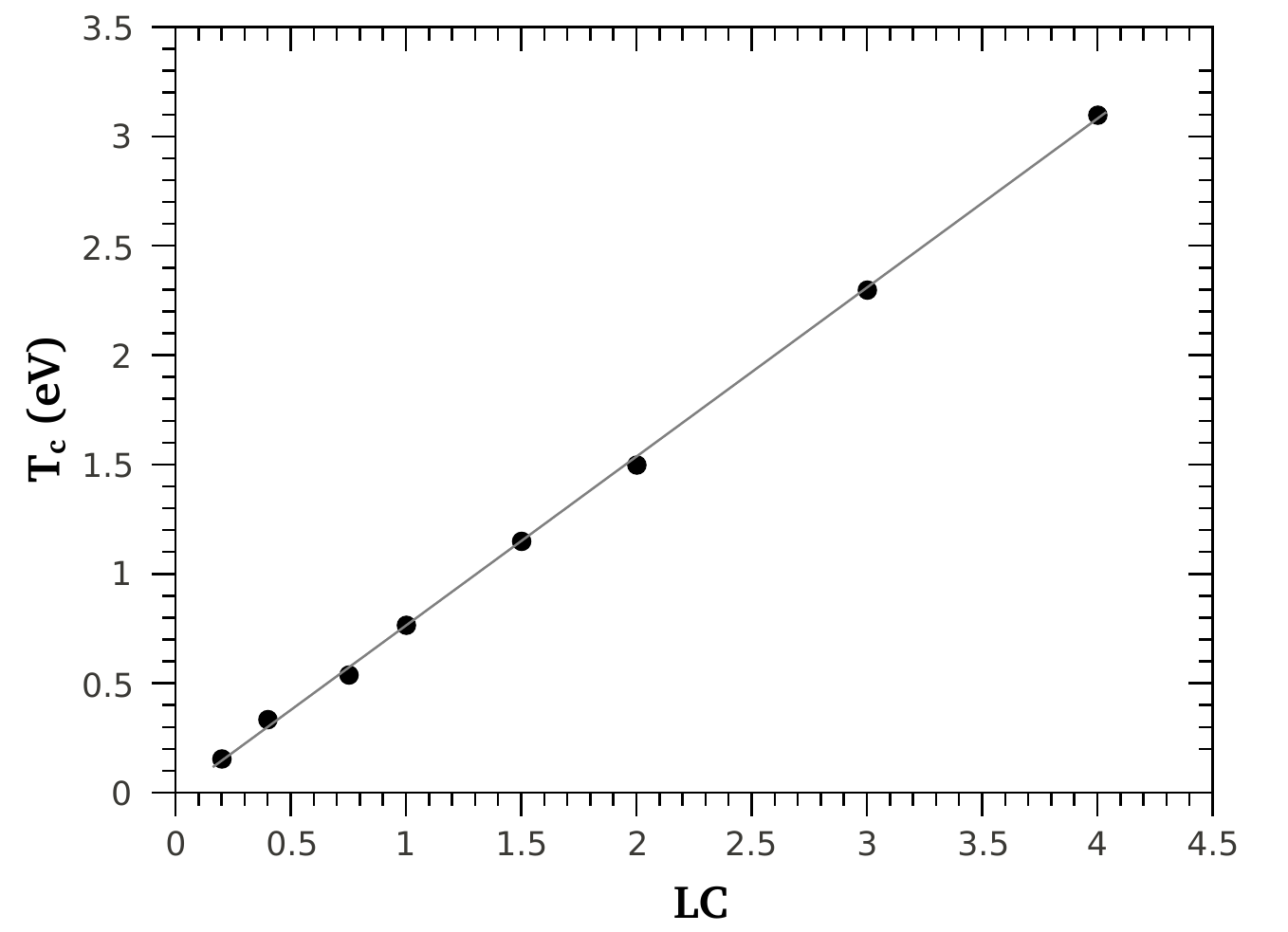}
    \else
      \includegraphics[bb = 92 86 545 742, height=2.2in]{Temperature-vs-LC}
    \fi
    \caption{Critical temperature against LC.}
    \label{TatBC05-vs-L}
  \end{center}
\end{figure}

In our case the Binder cumulant curves do not cross at a given point (see Fig. \ref{Binder-cumulant-vs-T-for-C}) which
makes it impossible to define a critical temperature. 
However, we have realized that if we plot the Binder cumulant  against the temperature for each value of the product
of the size and the concentration $LC$, then 
we obtain a crossing point (see Fig. \ref{Binder-cumulant-vs-T-for-LC}).  From this we obtain a 
relation between the Curie temperature $T_{\rm C}$ and $LC$ (see Fig. \ref{TatBC05-vs-L}):

\begin{equation}
   T_{\rm C}= (0.77\pm0.01) LC \hspace{10 mm} ({\rm eV})
\label{Curie_Temp}
\end{equation}

% -------------------------- Binder-cumulant-vs-T-for-LC ---------------------------------------

% ------------------------------------------------------------------------

% -------------------------- TatBC05-vs-L ---------------------------------------

% ------------------------------------------------------------------------

Strictly speaking the concept of Curie temperature should be used with caution since the ordering temperature in the thermodynamic limit is not well-defined in this model. However, our numerical simulations show clearly a measurable ordering temperature in {\em any} finite lattice. The expression (\ref{Curie_Temp}) and the Binder cumulant analysis of Figs. \ref{Binder-cumulant-vs-T-for-LC}, and  \ref{TatBC05-vs-L} admit a simple interpretation: If the system
is going to have  a well-defined critical temperature in the thermodynamic
limit and the Binder cumulant analysis is going to be an accurate method to determine it,
the coupling constant has to be rescaled with the size of system
\begin{equation}
  J'_0(LC) =\frac{J_0}{LC},
\label{rescaling_coupling}
\end{equation}
which redefines a coupling constant with units of energy. 
Without such rescaling the Binder cumulant analysis results in no crossing points 
(see Fig. \ref{Binder-cumulant-vs-T-for-C}).

This behavior is very common in systems with long-range couplings.
A very illustrating example  is the infinite-range Ising model (see for instance \cite{binney1992theory}),
where the coupling constant has to be rescaled with the total number of spins to achieve a well-defined 
critical temperature in the thermodynamic limit.
In our model an equally simple scaling argument can be offered to justify the re-scaling implicit in
Eq. (\ref{rescaling_coupling}). The effective coupling of a single spin connected by a $1/r$ 
interactions to the rest of the spins in the system is
\begin{equation}
  \langle J \rangle  \sim \frac{C}{a^2} \int_{0}^{La} \frac{J_0 a}{r}  r dr d\theta \sim  {J_0} LC .
\end{equation}
In other words, the effective coupling of the system increases linearly as its size increases. This is in contrast with a system with a finite coordination number where  $\langle J \rangle$ is size independent.  
Here we have assumed the continuum limit, a circular sample, and we have replaced the stochastic variable 
$p_j$ by its mean value $C$. We can remove these assumptions by evaluating numerically 
the effective coupling  $\langle J \rangle$ in the triangular discrete lattice with a random 
population of hydrogen atoms distributed with the probability density (\ref{probability_density}):
\begin{equation}
\langle  J_i  \rangle   =  \sum_{j}  J_{ij} p_{j},
\end{equation}
which, averaged over all the sites $i$ of the lattice, also scales as $LC$ in the limit of large 
cell size $\frac{L}{\ell} \gg 1$.

\begin{figure}[!htbp]
  \begin{center}
    \leavevmode
    \ifpdf
      \includegraphics[height=2.2in]{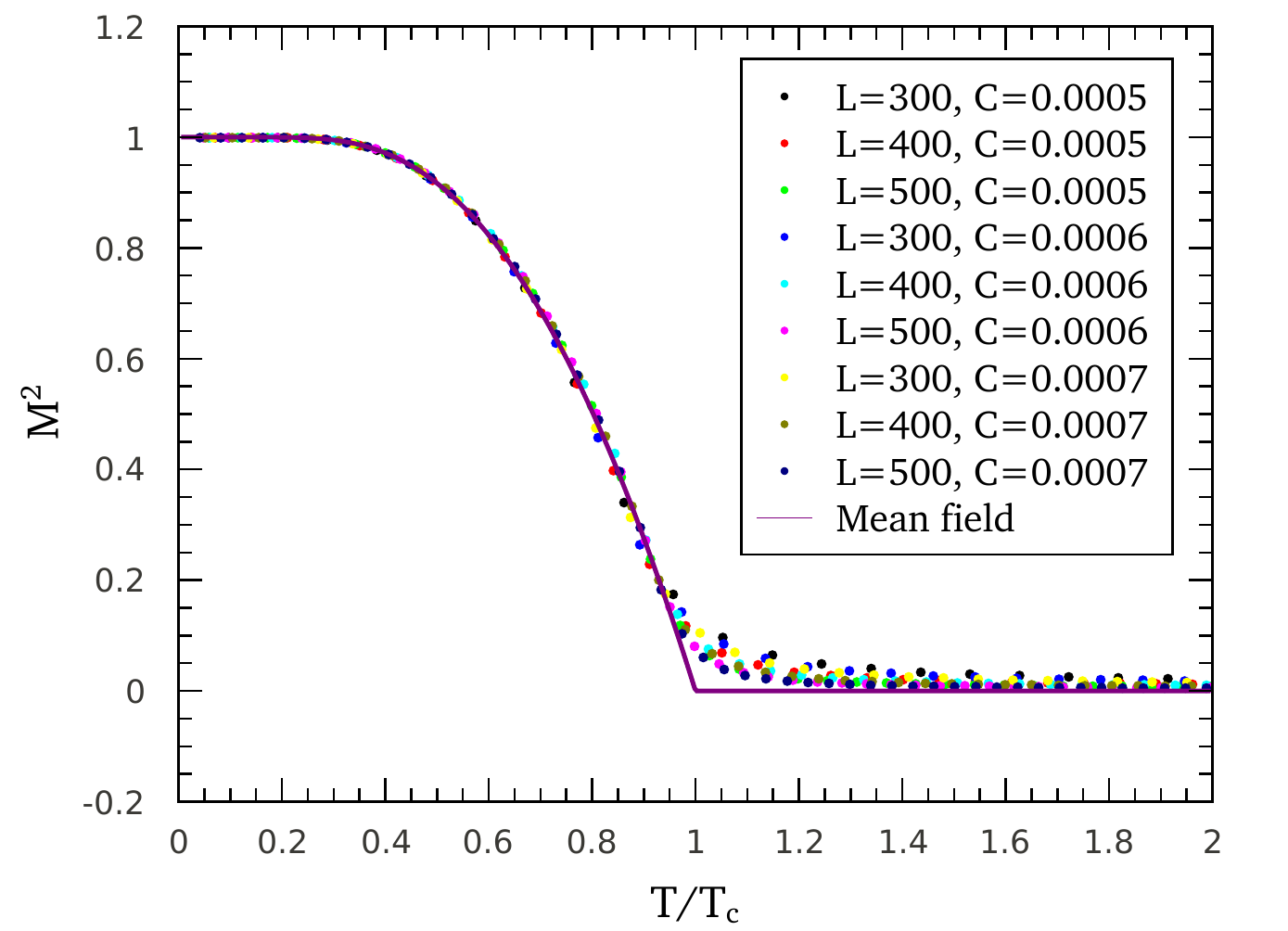}
    \else
      \includegraphics[bb = 92 86 545 742, height=2.2in]{M2-vs-T-over-Tc-for-MC-and-MF-0}
    \fi
    \caption{(Color online). Magnetization square vs. temperature over critical temperature 
computed with Monte Carlo and compared with the  mean-field result for various concentrations and sizes.}
    \label{M2-vs-T-over-Tc-for-MC-and-MF-0}
  \end{center}
\end{figure}

Finally we have compared the MC simulations with the mean-field approximation  (see
Fig. \ref{M2-vs-T-over-Tc-for-MC-and-MF-0}). In ordered long-ranged/high-coordination systems this approximation
can even be exact (see Ref. \onlinecite{Mean-Field} and references therein). In our model the agreement is very good.

%In Figure (\ref{M2-vs-T-over-Tc-for-MC-and-MF-0}) we show such an analysis for our model. While the qualitative behavior of $U_L(T)$ shows ordering for each independent supercell size, the ordering temperature $T_{O}(L)$ is clearly size dependent. This behavior is related with the long-ranged nature of the couplings. To illustrate this point, we computed  the effective coupling  between a single magnetic moment and the rest of the system in the mean field approximation.

%to do that, we rewrite the Hamiltonian in the form:

%\begin{equation}
%H=\sum_{i} S_{i}  p_{i}  \sum_{j}  J_{ij} p_{j} S_{j}
%\end{equation}
%in the mean-field approximation  $S_j$  is substituted by its site-dependent mean field value
%$\langle S \rangle$ :

%\begin{equation}
%H=\sum_{i} S_{i}  p_{i}   \langle S_j \rangle \sum_{j}  J_{ij} p_{j}= \sum_{i} \bar{J}_{i}S_{i}  p_{i}   \langle S \rangle
%\end{equation}

%the expression above defines a coupling $J_{i}$ as :
%\begin{equation}
% \ J_i=  \sum_{j}  J_{ij} p_{j}
%\end{equation}

%When averaged over all lattice sites in the limit of large supercell size $\frac{L}{\ell} \gg 1$ $\bar{J_i}$ depends on the size and concentration as $  =0.358CL$.

% -------------------------- Monte Carlo, and mean field method ---------------------------------------
% ------------------------------------------------------------------------

\vspace{50 mm} % ------------------------------------------------------------------------

% -------------------------------- Conclusion -----------------------------------

\section{Conclusions}
Through extensive DFT calculations we have found that
that the interaction between H atoms on graphene favors adsorption on different
sublattices along with an antiferromagnetic coupling of the induced magnetic moments. On the contrary, when hydrogenation takes place
on the surface of graphite or graphene multilayers (in Bernal stacking), the difference in 
adsorption energies takes over the interaction between H atoms and may result in all atoms adsorbed on 
the same sublattice and, thereby,  in a ferromagnetic state for low concentrations. Based on the exchange couplings
obtained from the DFT calculations, we have also evaluated the Curie temperature by mapping this system onto an Ising-like model with randomly located spins. The long-range nature of the magnetic coupling makes the Curie temperature size dependent
and larger than room temperature for typical concentrations and sizes.

% -------------------------------- Acknowledgment ---------------------------

\section{Acknowledgments}

This work was supported by MICINN under Grants Nos.
FIS2010-21883, CONSOLIDER CSD2007-0010, F1S2009-12721, FIS2012-37549,
 and by Generalitat Valenciana under Grant PROMETEO/2012/011. We also acknowledge computational support 
from the CCC of the Universidad Aut\'onoma de Madrid. We thank F. Yndur\'ain for discussions and 
I. Brihuega for sharing with us preliminary experimental data.

% ---------------------------- References -------------------------------------

%\bibliographystyle{apsrev}
%\bibliography{matcon,references} % References file

% ---------------------------------- END --------------------------------------

\end{document}